\documentclass[aps,pre,preprint,groupedaddress,nofootinbib]{revtex4-1}

\usepackage{bbm}
\usepackage{amssymb}
\usepackage{graphicx}
\usepackage{amsmath}
\usepackage{amsfonts,amsthm,bm} 
\usepackage{mathtools}
\usepackage{times}
\usepackage{bm}
\usepackage{natbib}
\usepackage{url}
\usepackage{multirow}
\usepackage{tabularx, ragged2e, booktabs}
\newcolumntype{Y}{>{\centering\arraybackslash}X}
\usepackage{adjustbox}
\usepackage{xcolor}

\usepackage{cprotect} 

\usepackage[update,prepend]{epstopdf} 

\begin{document}

\title[Global Stability of Climate: Melancholia States, Invariant Measures, and Phase Transitions]{Global Stability Properties of the Climate: Melancholia States, Invariant Measures, and Phase Transitions}

\author{Valerio Lucarini$^{1,2,3}$ and Tam\'as B\'odai$^{1,2,4,5}$}
\affiliation{$^1$Department of Mathematics and Statistics, University of Reading, Reading UK\\
$^2$Centre for the Mathematics of Planet Earth, University of Reading, Reading UK\\\
$^3$CEN, University of Hamburg, Hamburg, Germany\\
$^4$ Pusan National University, Busan, Republic of Korea\\
$^5$ Center for Climate Physics, Institute for Basic Science, Busan, Republic of Korea}
\date\today
\begin{abstract}
{ 
For a wide range of values of the incoming solar radiation, the Earth features at least two attracting states, which correspond to competing climates. The warm climate is analogous to the present one; the snowball climate features global glaciation and conditions that can hardly support life forms. Paleoclimatic evidences suggest that in the past our planet flipped between these two states. The main physical mechanism responsible for  such an instability is the ice-albedo feedback. In a previous work, we defined the Melancholia states that sit between the two  climates. Such states are embedded in the boundaries between the two basins of attraction and feature extensive glaciation down to relatively low latitudes. Here, we explore the global stability properties of the system by introducing random perturbations as modulations to the intensity of the incoming solar radiation. We observe noise-induced transitions between the competing basins of attraction. In the weak noise limit, large deviation laws define the invariant measure, the statistics of escape times, {\color{black}and  typical escape paths called instantons. }By constructing the instantons empirically, we show that the Melancholia states are the gateways for the noise-induced transitions. In the region of multistability, in the zero-noise limit, the measure is supported only on one of the competing attractors. For low (high) values of the solar irradiance, the limit measure is the snowball (warm) climate. The changeover between the two regimes corresponds to a first-order phase transition in the system. The framework we propose seems of general relevance for the study of complex multistable systems. 
Finally, we {\color{black}put forward} a new method for constructing Melancholia states from direct numerical simulations, {\color{black}which provides a possible alternative with respect to the edge-tracking algorithm.}}
\end{abstract}
\maketitle
\section{Introduction}\label{intro}
In the late 1960s and in the 1970s, Budyko, Selllers, and Ghil \cite{Budyko1969,Sellers1969,Ghil1976} proposed the idea that the Earth, in the current astrophysical and astronomical configuration, supports two co-existing attractors, the warm (W) state we live in, and the so-called snowball (SB) state, which is characterised by global glaciation and globally averaged surface temperature of about 200-220 K. Using parsimonious yet physically meaningful energy balance models, they indicated that the bistability of the climate system is the result of the competition between the positive ice-albedo feedback (a glaciated surface reflects  the incoming radiation more effectively) and the negative Boltzmann radiative feedback (a warmer surface emits more radiation to space). The relevance of the theoretical \textit{ansatz} became apparent when paleoclimatic data showed that, indeed, our planet has been flipping in and out of states of global glaciations corresponding to the predicated SB states during the Proterozoic, about 650 Mya \cite{Kirschvink,Hoffman2002,Pierrehumbert2011a}. 
W 
According to these energy balance models, the Earth's climate is bistable for a substantial range of values of the solar irradiance $S^*$, which include the present day value. Below the critical value $S^*_{W\rightarrow SB}$, only the SB state is permitted, whereas above the critical value $S^*_{SB\rightarrow W}$, only the W state is permitted.  Such critical values, which determine the boundaries of the region in parametric space where bistability is realised, are defined by bifurcations that occur when, roughly speaking, the strength of the positive, destabilising feedbacks becomes as strong as the negative, stabilizing feedbacks. Indeed, models of different levels of complexity ranging up to the state-of-the-art Earth System Models currently used for climate projections agree on predicting the existence of multistability in the climate system and point to the fundamental mechanisms described above as responsible for it, as well as providing values for $S^*_{W\rightarrow SB}$ that are in broad agreement with those obtained using simple models  \cite{Hyde2000,Voigt2010,Lucarini2010a}. 
We remark that both the concentration of greenhouse gases and the position of the continents have an impact on the values of  $S^*_{W\rightarrow SB}$ and $S^*_{SB\rightarrow W}$ and on the properties of the W and SB states \cite{Boschi2013}. Extremely high values of the concentration of CO$_2$ seem to be needed to deglaciate from a SB state \cite{crowley2001}. 

Improving our understanding of the critical transitions associated to such a bistability is a key challenge of geosciences and has strong implications also in terms of the quest for understanding or anticipating planetary habitability. Planets in the habitable zone have astronomical and astrophysical configurations that allow, in principle, the presence of water at surface. Therefore, $S^*_{W\rightarrow SB}$ defines the cold boundary of the habitable zone. Clearly, an exoplanet in the habitable zone can be in the regime of bistability: if the planet is in the SB state, it will have very hard time supporting life\footnote{The project EDEN (http://project-eden.space/) combines  ideas and methods in astrophysics, planetary sciences, and astrobiology to search for and characterize nearby habitable worlds.}. Additionally, astronomical parameters such as the obliquity of the planet \cite{Linsenmeier2015,Kilic2017,Kilic2018}, {\color{black}eccentricity \cite{Linsenmeier2015},} or the length of the year \cite{LucAstr2013,Abbot2018} can have a dramatic effect on the properties of multistability of a planet in the habitable zone, up to the point of erasing it altogether. {\color{black}In particular, planets with Earth-like atmospheres and high seasonal variability can have ice-free areas at much larger distance from the host star than planets without seasonal variability, which leads to a substantial expansion of the outer edge of the habitable zone. Additionally}, one expects that tidally locked planets with an active carbon cycle can never be found in a SB state \cite{Checlair2017}.

{\color{black}\subsection{Multistability of the Climate System}}
Further investigations have proposed the possibility of the existence of alternative cold states with respect to the SB one. Such states are characterised by the existence of a thin strip of ice-free region near the equator. Clearly, this possibility has key implications in terms of habitability and the evolution of life on Earth. Different physical mechanisms have been proposed for explaining the existence of such a state, based either on the role of a dynamical ocean \cite{lewis2007} or the specific properties of the albedo of sea ice \cite{abbott2011}.  In fact, in a previous work \cite{Lucarini2017} we have found that, indeed, a third co-existing stable  state, intermediate between the SB and the W climate, can be found in a climate model featuring a very simplified  representation of the oceanic heat transport, so that one might expect that the {\color{black}existence of more than two competing attractors could be a rather robust  }property of the climate system.  

{\color{black}Indeed, the dynamical landscape of the Earth system might be even more complex than what is usually expected. A recent study  \cite{brunetti2019}, performed using a rather sophisticated climate model run using aquaplanet boundary conditions (without continents), indicates the existence of at least five competing climate states,  ranging from a snowball to a very warm state without sea ice. 

Additionally, the physics and the chemistry of the climate system feature further complexities when one considers even warmer conditions. For sufficiently large values of $S^*$, the W climate state loses its stability as a result of the dramatic strengthening of the positive feedback associated to the presence of water vapour in the atmosphere. Indeed, warm conditions favour, through the thermodynamic effect associated with the Clausius-Clayperon relation, the retaining capacity of water vapour of the atmosphere. The water vapour is a powerful greenhouse gas, as it is active in the infrared radiation. As a result, when the concentration of water vapour is sufficiently high, the planet performs a transition to either the so-called moist greenhouse state or the so-called  runaway greenhouse state (associated to a complete evaporation of the oceans) \cite{GomezLeal2018a,GomezLeal2018b}, which define the warm (or inner) edge of the habitable zone \cite{Kastings1993}.}


In what follows, we consider the  simpler - yet extremely relevant -  scenario where the only relevant co-existing climates are the SB state and the W state. As discussed in \cite{Lucarini2010a}, the physics of the system is especially interesting when the critical transitions are approached. As the solar irradiance  $S^*$  nears the critical value $S^*_{W\rightarrow SB}$ with the system being in the W state, the climatic engine becomes more efficient, because larger temperature gradients are realised inside the domain. Such an increased efficiency leads to a stronger atmospheric circulation, which is fuelled by temperature gradients and tends to reduce them by transporting heat from warm to cold regions, acting as a non-trivial diffusion process. Such a non-linear equilibration mechanism acts as a negative feedback and, broadly speaking, is a macroscopic manifestation of the second law of thermodynamics. One of the results of the heat transport performed by the atmospheric circulation is the stabilization of the ice-line. When $S^*=S^*_{W\rightarrow SB}$, the ice-albedo feedback becomes as strong as the negative feedbacks of the system, and the system flips to SB state with the ice-line reaching the equator. Similar mechanisms are in place when the system is in the SB state and $S^*$  nears, instead, $S^*_{SB\rightarrow W}$. When $S^*=S^*_{SB\rightarrow W}$, the ice begins to melt near the equator, leading to a rapid poleward retreat of the ice line. 

At the critical transitions the climate system is not anymore able to dampen fluctuations due to (infinitesimal) external forcings. Using methods borrowed from  transfer operator theory \cite{Baladi2000}, the investigation of the behaviour of the same model used in \cite{Lucarini2010a}  has indeed shown that when $S^*$ nears $S^*_{W\rightarrow SB}$, the spectral gap - {\color{black}defined as the absolute value of the (negative) imaginary part of the subdominant Ruelle-Pollicott pole \cite{Pollicott1985,Ruelle1986} - of the transfer operator constructed in a suitably defined reduced phase space vanishes. As a result, exponential decay of correlation is lost and the system experiences what is often referred to as \textit{critical slowing down} \cite{Tantet2018}; see also Refs. \cite{Shiino1987,Pavliotis2014}.}

{\color{black}Far from critical transitions, it has been shown \cite{Ragone2016,Lucarini2017b,Lembo2019} that it is possible to perform climate change projections resulting from a time-dependent CO$_2$ forcing using Ruelle response theory \cite{Ruelle2009}. In the case of  perturbations not depending explicitly on time, response theory allows one to describe how the measure of the system changes differentiably with respect to small changes in the dynamics of the system. In the case of time-dependent perturbations, response theory makes it possible to reconstruct the measure supported on the pullback attractor \cite{Ghil2008,Chekroun2011,CLR13} {\color{black}(see also the closely related concept of snapshot attractor \cite{Ott.ea.1990,DBT15})} the non-autonomous system through a perturbative approach around a reference state, which, in the case of the climate studies referred to here, corresponds to the pre-industrial conditions. When we are nearing a critical transition, it is reasonable to expect a monotonic decrease of the spectral gap \cite{Dembo2010,Assaraf2018}. Hence, in the vicinity of the critical transitions the  presence of a vanishing spectral gap leads to having a vanishing radius of expansion for  response theory  \cite{Lucarini2016}. Indeed, it is expected that the (near) closure of the spectral gap is associated to a strongly enhanced sensitivity of the system's statistics to perturbations \cite{Chekroun2014}.  See \cite{Ghil2019} for a thorough discussion on the various regimes of climatic response to forcings and of the relationship between climate change and climate variability across various temporal scales.
}

\begin{figure}
a) \includegraphics[trim=1cm 0cm 0cm 1.22cm, clip=true, width=0.7\textwidth]{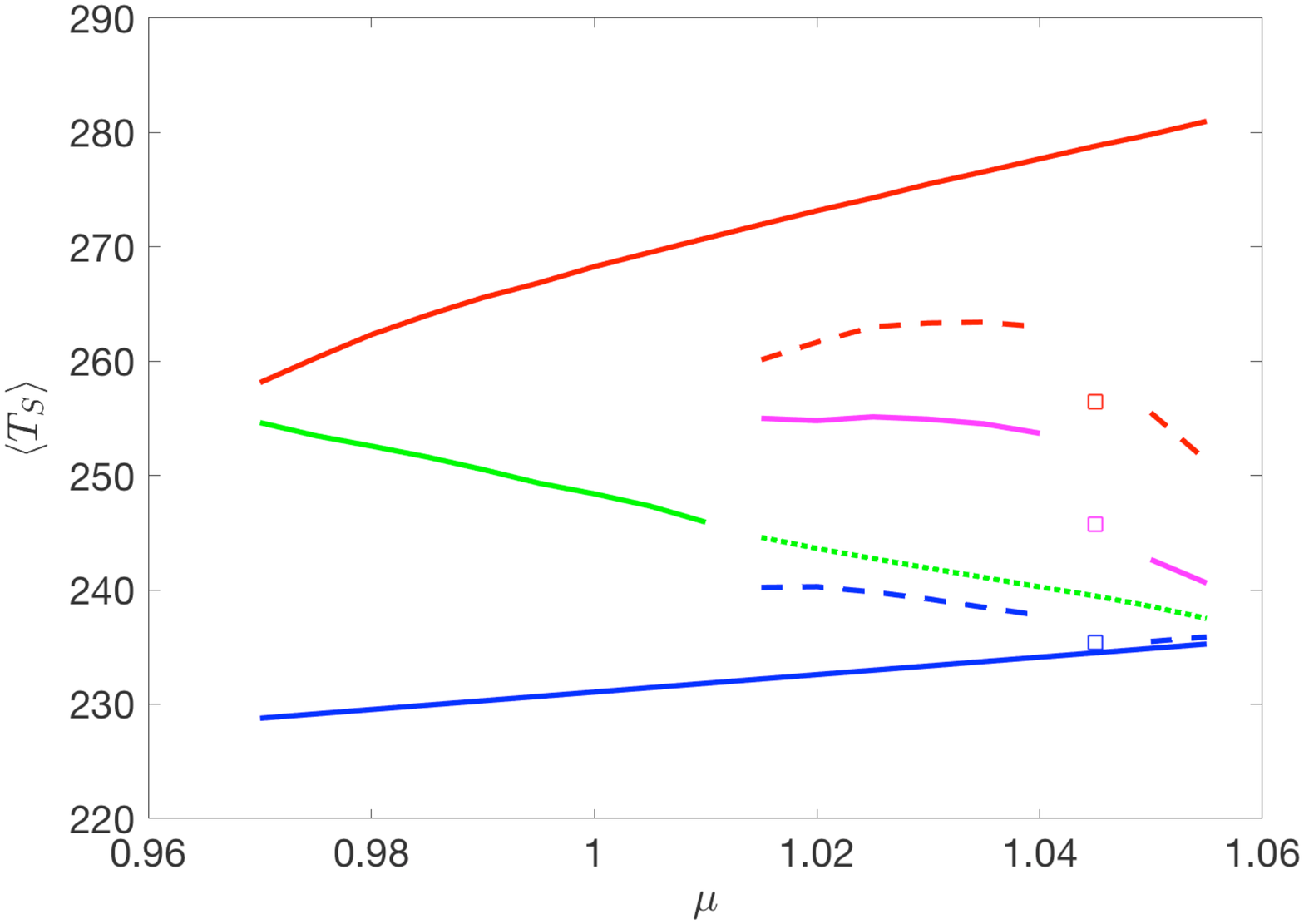}\\
b) \includegraphics[trim=1cm 0cm 0cm 1.22cm, clip=true, width=0.7\textwidth]{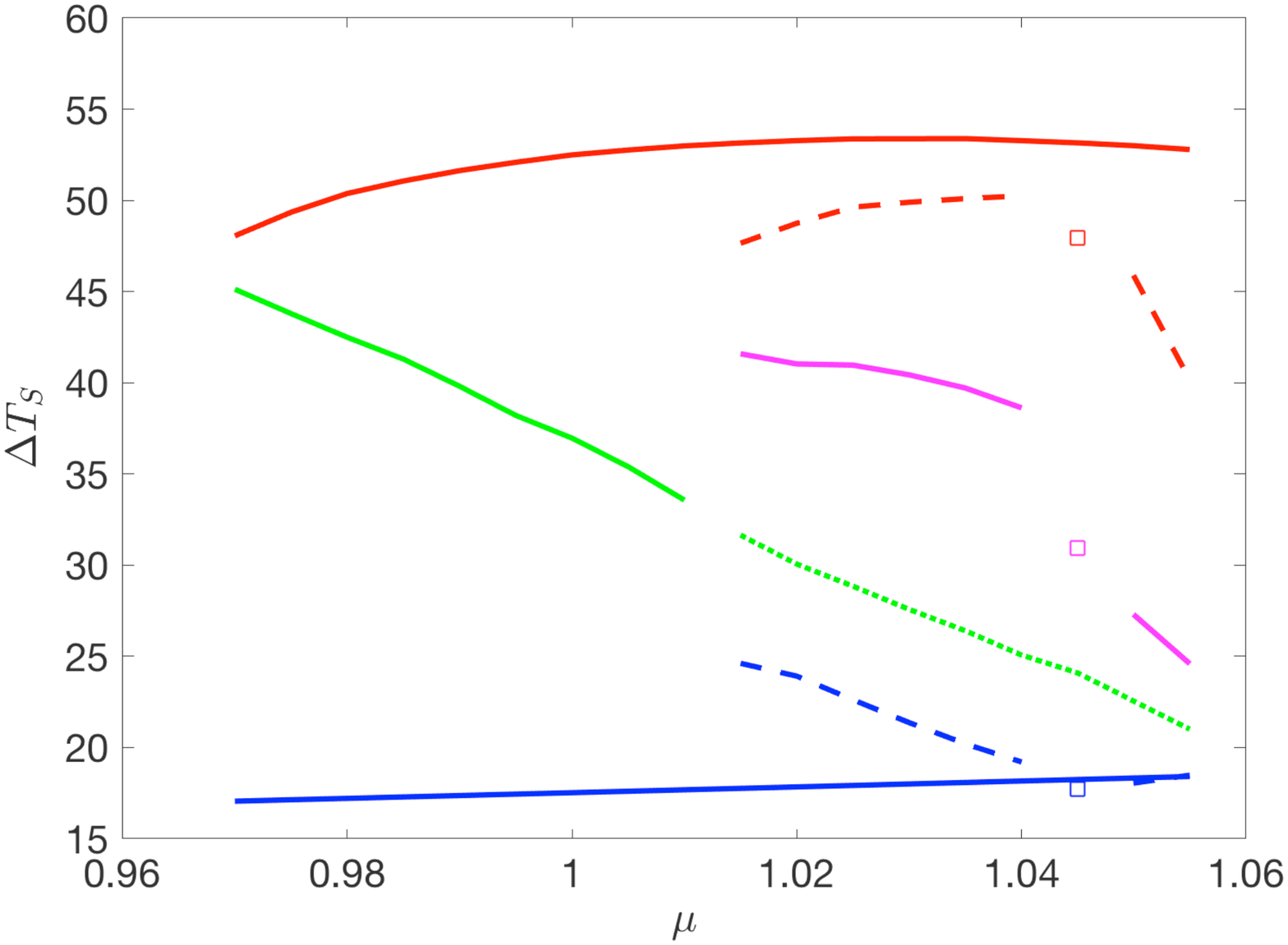}\
\caption{Bifurcation diagram for the coupled climate model studied in \cite{Lucarini2017}. Panel a) Globally averaged ocean temperature $\langle T_S \rangle $ vs  $\mu$. Bistability is found for a large range of values of $\mu$. Red continuous line: W attractor; blue continuous line: SB attractor; green continuous line: chaotic M state; purple continuous line: mean properties of the symmetry broken chaotic M state: red dashed line: warm side of the symmetry broken chaotic M state; blue dashed line: cold side of the symmetry broken chaotic M state; green dotted line; transient symmetric chaotic M state; empty squares: warm side (red), cold side (blue) and average properties of the third attractor. The W-to-SB tipping point is located at $\mu\sim 0.965$; the  SB-to-W tipping point is located at $\mu\sim 1.055$;. See \cite{Lucarini2017}. \label{oldbifurcation}}
\end{figure}

{\color{black}\subsection{Melancholia States of the Climate System}}
 A key question is what lies in-between the stable co-existing climates within the region of the parameters space where bistability is found. In simple models, it is often possible to identify unstable solutions sitting in-between the two stable climates. {\color{black}Such unstable solutions are embedded in the boundary between the two basins of attraction and, are roughly speaking, ice-covered up to the mid-latitudes. These solutions are saddles because they attract orbits starting from initial conditions on the basin boundary but are not stable, as small generic perturbations pushing the orbit outside the basin boundary with probability one and then lead to} the system falling eventually into one of the competing attractors \cite{Ghil1976,Bodai2014}. 
         
        
         When studying more comprehensive climate models featuring chaotic dynamics, things are, as described in the next section, considerably more complex from a mathematical point of view, and the individuation of the unstable saddle solutions is much harder \cite{Grebogi1983,Robert2000,Ott2002,LT:2011}. Since these solutions are unstable, they cannot be found by direct numerical simulation. {\color{black} In a previous investigation \cite{Lucarini2017}, we adapted the edge tracking algorithm \cite{Skufca2006,Schneider2007} introduced for constructing the edge states, i.e. the special solutions separating  laminar from long-lived turbulent regimes of motion in a fluid dynamical setting\footnote{Note that in such systems, technically, there are no competing attractors, because the only true attractor is the laminar state.}. We used a recursive technique of bisection on the initial conditions\footnote{A different approach based on control theory aims at finding unstable solutions by a feedback loop involving changes in the value of the control parameter defining the region of bistability \cite{Sieber2013,Sieber2014}. } for shadowing trajectories on the basin boundary separating the two co-existing W and SB states, and managed to populate the corresponding saddles.      
         
         The analysis was performed using an intermediate complexity climate model with $O(10^4)$ degrees of freedom. The saddles we found had the remarkable property of featuring chaos {\color{black}(we found evidence that the first Lyapunov exponent was positive)}, and were named as Melancholia (M) states.} Figures \ref{oldbifurcation}a)-b) summarize the main properties of the system, by showing the long term averages of the globally averaged surface temperature $\langle T_S \rangle$ and of the temperature difference between low and high latitudes $\Delta T_S$ as a function of the relative solar irradiance $\mu=S^*/S^*_0$, where $S^*_0$ is the present day value. By focusing on the M states we have been able to show the existence of much richer than previously thought dynamical landscape.  
         
{\color{black}As discussed in \cite{Lucarini2017}}, up to  $\mu\sim 1.01$, the M state is characterised by longitudinal symmetry in its statistical properties, just as the boundary conditions of the system, are, indeed, longitudinally symmetric. The chaotic dynamics manifests itself as weather variability in a form not too dissimilar from the usual one observed in stable climates. Nonetheless, on long time scales,  orbits initialised near the M states drift to either the W or the cold SB state, as a result of the dominating positive ice-albedo feedback. For $\mu\sim 1.01$, we observed that the symmetric M state  becomes transient, evolving very slowly (on a time scale scale much longer than the other ones typical of the system) into a symmetry-broken state, where very cold and very warm conditions co-exist, separated by two regions of very strong \textit{longitudinal} temperature gradient. The two regions feature rather different dynamical behaviour and the boundary between them rotates very slowly in time. The nontrivial bifurcation associated to such a symmetry break  leads to dynamical regimes that resembles \textit{chimera states} in extensive systems \cite{Abrams2004,Omelchenko2018Nonlinearity}. The third climate state mentioned before exists in a small parametric window near $\mu\sim 1.045$ \cite{Lucarini2017}. 

We remark that the dynamical systems viewpoint clarifies that the critical transitions for the W (snowball) state occurring when $S^*$ approaches the critical value $S^*_{W\rightarrow SB}$ ($S^*_{SB\rightarrow W}$) are associated to the collision between the M state and one of the W (snowball) climates, according to the dynamical scenario of boundary crisis \cite{Ott2002}. The system's reduced ability to dampen fluctuations near the tipping points and the associated shrinking of the spectral gap described above can be seen, dynamically, as the result of the fact that the attractor \textit{attracts orbits less effectively}  in its immediate neighbourhood because of the presence of a nearby M state state \cite{Tantet2018}.

\subsection{This Paper: Goals and Main Results}
Studying multistable systems in general, and the climate system in particular, using deterministic autonomous dynamical systems faces two important issues, both resulting from the fact that the phase space is partitioned into disjoint invariant sets - the various basins of attractions and their respective boundaries. First, it is not possible to account for transitions between the co-existing basins of attraction. Instead, transitions between distinct \textit{regimes of motion} are  observed in many systems of interest. Additionally, one cannot establish an ergodic, physically relevant invariant measure, as the co-existing attractors are disjoint. {\color{black}Assigning a weight to each of them is, indeed, a highly arbitrary operation}. Hence, one cannot answer the question of how likely it is for the system of interest to be found at a given time in a specific regime of motion. Instead, all the statistical properties of an orbit are conditional on which invariant set its initial conditions belong to. Indeed, building upon the preliminary results reported in a short communication \cite{LB2018} and giving them a much broader scope and setting them a in much more robust mathematical framework, we want to address these points. For the benefit of the reader, we report below the main goals of our investigation and the main findings presented in this paper. 

We will introduce a stochastic forcing  to the model studied in \cite{Lucarini2017} as a Gaussian perturbation with variance $\sigma^2$ modulating the intensity of the incoming solar radiation. This impacts the radiative budget of the planet in a very nontrivial way and, thanks to the ice-albedo feedback, acts as a multiplicative noise. Additionally, we conjecture, supported by {\color{black}physical  and qualitative mathematical arguments}, that the combination of noise and nonlinear deterministic dynamics leads to a hypoelliptic diffusion process, i.e. the noise propagates to all degrees of freedom of the system \cite{Bell2004}. The presence of a random forcing allows the system to perform transitions between the neighbourhoods of the deterministic attractors, crossing the basin boundaries that, in the unperturbed case, are impenetrable; see some classical results in \cite{Hanggi1986,Kautz1987,Grassberger1989}. The consideration of a random forcing will allow us to construct the ergodic invariant measure of the system  by performing long integrations, and investigate the properties of noise-induced transitions. 
         
         Following \cite{Freidlin1984,Graham1991,Hamm1994,LT:2011}, {\color{black}we  show that, in the weak noise limit and under suitable conditions, the invariant measure can be written as a large deviation law with the following notable properties. The exponent is given by minus the quasi-potential function divided by $\sigma^2/2$. {\color{black}  We will show how to compute the quasi-potential from the drift and volatility fields, and show that the drift field can be decomposed in two contributions having radically different dynamical meaning.  Additionally, the quasi-potential is a Lyapunov function and provides a clear picture of the evolution of the system in the absence of noise. } In the region of bistability, the quasi-potential has local minima associated to the attractors and has a saddle behavior at the M states of the deterministic system. } In the region where only one stable state is realised in the deterministic case, the quasi-potential has just one local (and global) minimum, corresponding to the unique attractor.  
         
In the case of multistability, the logarithm of the average permanence time in a basin of attraction increases linearly with the product of $2/\sigma^2$ times the difference between the value of the quasi-potential at the M state through which the transitions takes place and its value at the corresponding attractor. We also show that the stochastically averaged exit trajectories connect the attractor with the M state, which is indeed the most likely exit point from the deterministic basin of attraction. In the weak noise limit, such paths correspond to the instantons \cite{Kautz1987,Grassberger1989,Kraut2002,Beri2005,Bouchet2016}. In our simulations, we show that such an identification becomes more accurate as the intensity of the noise is decreased. Exploiting this property, we also propose a new  method for constructing the M states via \textit{direct} numerical integration of stochastic differential equations. 
         
         We  discover that, since generally the local minima of the quasi-potential  corresponding to the two attractors have different values, in the zero-noise limit only one attractor - the one corresponding to the global minimum of the quasi-potential - is populated. Nonetheless, an individual trajectory may in fact persist near the competing metastable state for a very long time, as the permanence time in the corresponding basin of attraction also diverges (but at a slower rate than for the asymptotic state). As indicated by physical intuition, for low values of $S^*$, the noise selects as limit measure the SB state, while for high values of $S^*$, the limit measure is the W attractor. The changeover takes place for a critical value of the {\color{black}relative solar irradiance $\mu=\mu_{crit}$, where  $\mu_{crit}\approx 1.005$. For such a value of $\mu$}, the equivalent of a first order phase transition for equilibrium systems takes place in the system. The order parameter that more obviously captures the transition is the globally averaged surface temperature. 
   
                  
         The paper is structured  as follows. In Section \ref{Mathematics} we summarise the main mathematical concepts and ideas we have used to frame our investigation, to perform the data analysis, and to interpret our results. In Section \ref{ModelData} we report the modelling suite we have used, describe the data we have produced, and give information on how they have been processed. In Section \ref{Results} we present and discuss our results.  {\color{black}In Section \ref{appa} we propose and test numerically a new method for constructing the M states by looking at the intersections of instantons constructed by taking into account three different noise laws superimposed on to the same deterministic dynamics. In Section \ref{Conclusions} we summarize and interpret our findings, describe the limitations of our work, and propose ideas for future investigations. In Appendix \ref{appb}, we speculate on the fact that the framework presented in the paper is potentially suitable for clarifying the role played in complex historical processes by \textit{contingency}, as discussed by Gould \cite{G1989} in the context of evolutionary biology.}


\section{Mathematical Background}\label{Mathematics}
\subsection{Geometry of the Phase Space: Attractors and M States}
The investigation of systems possessing multiple steady states is an extremely active area of interdisciplinary research, encompassing mathematics, natural sciences, and  social sciences. The recent review by Feudel et al. \cite{Feudel2018}, introducing a special journal issue on the topic, gives a rather complete overview of the state-of-the-art of the ongoing activities on this topic and provides several interesting examples. In the context of Earth sciences, it is more common to refer to critical transitions in multistable systems using the expression  \textit{tipping points}, which has received a great popularity following a paper by Lenton \textit{et al.}  \cite{Lenton2008}.

A possible way to introduce the  mathematical background for multistable systems is the following. We consider a smooth autonomous continuous-time deterministic dynamical system defined on a smooth finite-dimensional compact manifold $\mathcal{M}$ (in what follows, a subset of $\mathbb{R}^N$). {\color{black}We assume that the dynamical system is dissipative, so that the phase space continuously contracts and the Lebesgue measure is not conserved.} The orbit evolves from an initial condition $\mathbf{x}_0\in\mathcal{M} $ at time $t=0$. We define $\mathbf{x}(t,\mathbf{x}_0)=S^t(\mathbf{x}_0)$ as the orbit at a generic time $t$, where $S^t$ indicates the evolution operator. {\color{black}The corresponding set of ordinary differential equations written componentwise is:
\begin{equation}\label{eq1}
{\dot{x}_i}={F}_i(\mathbf{x}), \quad i=1,\ldots,N,
\end{equation}}
where $\mathbf{F}(\mathbf{x})=\textrm{d}/\textrm{d}\tau S^\tau(\mathbf{x})|_{\tau=0}$ is a smooth $N$-dimensional vector field. We define such a dynamical system as multistable if it features more than one asymptotic states. Specifically, we mean that {\color{black}there are two or more attractors $\Omega_j$, $j=1,\ldots,J$, each a corresponding basin of attraction having a finite Lebesgue measure}. Each of the attractors is an invariant set and is the support of an invariant measure of the system. The asymptotic state where the trajectory falls into is determined by its initial conditions, and the phase space is partitioned between the basins of attraction $B_l$ of the various attractors $\Omega_j$ and the boundaries $\partial B_l$, $l=1,\ldots,L$ of such basins. 

We assume, for simplicity, that within each basin of attraction an ergodic measure  can be defined  as the limit of the empirical measure constructed by averaging over infinitely long forward trajectories {\color{black}for Lebesgue almost all initial conditions in the the basin of attraction}. In other terms, all the invariant measures of the system can be written as a convex sum of its $j=1,\ldots,J$ ergodic components, each supported on the corresponding attractor $\Omega_j, j=1,\ldots,J$.  

{\color{black}We also assume that orbits initialized on the basin boundaries $\partial B_l$, $l=1,\ldots,L$ are attracted towards invariant saddles. In each basin boundary $\partial B_l$ there can be one or more of such saddles. We assume that each saddle features only one unstable direction.
%
%
Such instability repels trajectories initialised near the saddle towards either of the competing attractors. 

{\color{black} In general, one can be in the situation where the asymptotic dynamics} on one or more of the competing attractors can be chaotic (meaning here that at least one Lyapunov exponent is positive and unstable periodic orbits are dense). In some systems, chaotic dynamics can be realised on one or more of the saddles  embedded in the basin boundaries \cite{Grebogi1983,Robert2000,Ott2002,Vollmer2009}. {\color{black}These are what we refer to as M states.} 
M states are non-trivial geometrical objects being the support of a typically nontrivial measure \cite{LT:2011}. 


A fascinating aspect of multistable systems is the following. If the first Lyapunov exponent of a chaotic saddle is
larger than the inverse of the life time of the saddle state itself (which measures the rate at which orbits initialised near the saddle state are repelled towards the two nearby asymptotic states), then the basin boundary separating the basin of attraction of the two asymptotic sets has co-dimension strictly smaller than one. In two-dimensional maps, it has been proved that in this case the basin boundary is not a manifold, but rather a \textit{rough fractal}  \cite{Grebogi1983,Vollmer2009,LT:2011}. The authors have recently proposed a generalisation of this result to the case of $N$-dimensional maps \cite{BodaiLucarini2020}. In \cite{Lucarini2017}, despite  high-dimensionality of the system, one could detect that the co-dimension  of the basin boundary is strictly smaller than one. In fact, such co-dimension was found to be very close to zero, as  a result of time scale difference between the most relevant instabilities: the climatic instability due to the ice-albedo feedback acting near the M state is much slower than the fast weather-like instability associated to baroclinic processes occurring on the M state. This implies that near the basin boundary there is basically no predictability of the second kind in the sense of Lorenz \cite{Lorenz1975}. The basin boundary is, \textit{de facto}, a grey zone, and it is difficult to assess where orbits initialised near the boundary will end up.

\subsection{Impact of Stochastic Perturbations: Invariant Measure and Noise-induced Transitions}\label{stochastic}

 The goal of our investigation here is to analyse the impact of imposing a random forcing on the deterministic dynamics discussed above. 
 Processes that can be described as a noise-induced escape from an attractor have long been studied in the natural sciences; see \cite{Hanggi1986,Kautz1987,Grassberger1989}.  
 {\color{black}
We generalise Eq. \ref{eq1} by adding a stochastic component. We then consider a stochastic differential equation (SDE) in It\^o form written as
\begin{equation}\label{eqapp}
{d{{x}}_i}={F}_i(\mathbf{x})dt+\sigma{s}(\mathbf{x})_{ij}d{W}_j,
\end{equation}
where $\mathbf{x},\mathbf{F} \in \mathbb{R}^{N}$, $\mathbf{F}$ and $s_{ij}\in \mathbb{R}^{N\times N}$ are smooth, $\sigma\geq 0 $, $d{W}_j$ is the increment of an $N-$dimensional brownian motion, and $C_{ij}(\mathbf{x})={s}_{ik}(\mathbf{x}){s}_{jk}(\mathbf{x})$ is the noise covariance matrix.  
We  consider the case of a hypoelliptic diffusion process. This amounts to assuming that while the covariance matrix of noise can be singular, the drift term modified according to the Stratonovich convention  and the columns of the matrix $s$ satisfy the so-called H\"ormander condition, i.e., the Lie algebra generated by them has dimension $N$ everywhere \cite{Hairer2011}. As a result of this, a smooth invariant density with respect to Lebesgue is realised because the noise is propagated to all the coordinates through the drift term; see  \cite{Bell2004}\footnote{Assuming an elliptic diffusion process is extremely restrictive because it requires  all degrees of freedom to be driven by Gaussian noise.}. We discuss below in Sect. \ref{oceansec} why such a mathematically important assumption can be heuristically justified in the specific case studied here. 

Taking inspiration from the Freidlin-Wentzell~\cite{Freidlin1984} theory and modifications thereof \cite{Graham1991,Hamm1994,LT:2011}, {\color{black} in the weak-noise limit $\sigma\rightarrow 0 $ we seek a special functional form for the invariant measure. Indeed, we look for } a large deviation law:
\begin{equation}\label{eq:stationary_distr}
  \Pi_\sigma(\mathbf{x}) \sim \exp\left(-\frac{2\Phi(\mathbf{x})}{\sigma^2}\right),
\end{equation}
where the rate function $\Phi(\mathbf{x})$ is referred to as quasi-potential, and we have neglected the pre-exponential term. Specifically, the symbol $\sim$ in Eq. \ref{eq:stationary_distr} implies that $\Phi(\mathbf{x})=-1/2\lim_{\sigma\rightarrow 0} \sigma^2  \log \Pi_\sigma(\mathbf{x})$.  The function $\Phi(\mathbf{x})$ can be obtained as follows. We solve the stationary Fokker-Planck equation corresponding to Eq. \ref{eqapp}:
\begin{equation}
\partial_j J_j(\mathbf{x})=0, \quad  J_j(\mathbf{x}) = -{F}_j(\mathbf{x})  \Pi_\sigma(\mathbf{x}) +\sigma^2\partial_i  \left(C_{ij}(\mathbf{x}) \Pi_\sigma(\mathbf{x}) \right),
\end{equation}
where $\mathbf{J}$ is the current density. We then consider the weak noise limit, and use as ansatz the expression given in Eq. \ref{eq:stationary_distr}. We obtain the following Hamilton-Jacobi equation \cite{Gaspard2002}:
\begin{equation}\label{eq:HJE}
{F}_i(\mathbf{x}) \partial_i \Phi(\mathbf{x})+C_{ij}(\mathbf{x})  \partial_i \Phi(\mathbf{x}) \partial_j \Phi(\mathbf{x}) =0.
\end{equation}
This equation allows one to express $\Phi$ in terms of the drift and volatility fields introduced in Eq. \ref{eqapp}. The quasi-potential $\Phi$ can also be computed by solving the variational problem associated with the Freidlin-Wentzell action \cite{Nardini2016}. Finally, alternative routes for computing $\Phi$ have been proposed in \cite{Ao2004,Yin2006}. 

The explicit computation of $\Phi$ is far from trivial, yet of great interest in many applications; see e.g., \cite{Zhou2012} for the case of biological systems. Brackston \textit{et al.} \cite{Brackston2018} have recently proposed an algorithm for estimating $\Phi$ in the case that the governing
equations are polynomial and involves solving an optimization over the coefficients of a polynomial  function. Instead, Tang \textit{et al.} \cite{Tang2017} proposed a variational method for estimating in the  populations corresponding to each deterministic attractor without resorting to computing the invariant measure.

 Following \cite{Graham1991,Hamm1994}, we now describe the dynamical meaning of $\Phi$. Indeed, solving the previous Hamilton-Jacobi equation corresponds to the fact that it is possible to write the drift vector field as the sum of two vector fields:
\begin{equation}\label{eq:decomposition}
{F}_i(\mathbf{x}) = {R}_i(\mathbf{x})- C_{ij}(\mathbf{x})\partial_j \Phi(\mathbf{x})
\end{equation}
that are mutually orthogonal, so that  ${R}_i(\mathbf{x})\partial_i \Phi(\mathbf{x})=0$. In the case Eq. \ref{eqapp} describes a thermodynamical system near equilibrium, $\mathbf{R}$ defines the time reversible dynamics, while $\mathbf{F}-\mathbf{R}$ defines the irreversible, dissipative dynamics \cite{Graham1987}.  
%
%
One finds that  
\begin{equation}\label{eq:Lyap}
d\Phi(\mathbf{x})/dt = -C_{ij}(\mathbf{x})  \partial_i \Phi(\mathbf{x}) \partial_j \Phi(\mathbf{x}) +  {R}_i(\mathbf{x})  \partial_i  \Phi(\mathbf{x}) = - C_{ij}(\mathbf{x})  \partial_i \Phi(\mathbf{x}) \partial_j \Phi(\mathbf{x}).
\end{equation}
As a result of this, $\Phi$ has the role of a Lyapunov function whose decrease describes the convergence of an orbit to the attractor. Specifically, $\Phi(\mathbf{x})$ has local minima at the deterministic attractors $\Omega_j$, $j=1,\ldots, J$, and has a saddle behaviour at the saddles  $\Pi_l$, $l=1,\ldots,L$. If an attractor (saddle) is chaotic, $\Phi$ has constant value over its support, which can then be a strange set \cite{Graham1991,Hamm1994}. 

Note that in the standard case of dynamics taking place in an energy landscape defined by a (confining) potential $U(\mathbf{x})$ and noise correlation matrix proportional to the identity (obtained by setting ${F}_i(\mathbf{x})=-\partial_i U(\mathbf{x})$ and $C_{ij}(\mathbf{x})=\mathbf{1}$ in the previous equations) one has $\Phi=U$. Additionally, one derives $\dot U(\mathbf{x})=-\partial_i U(\mathbf{x}) \partial_i U(\mathbf{x}) <0$ and  $U(\mathbf{x})$ is a Lyapunov function, and one recovers an equilibrium state, where detailed balance applies and, by definition, the current vanishes ($\mathbf{J}=0$)\footnote{Note that, in general, we can have an equilibrium state if and only if the the drift term has a gradient structure with respect to the metric defined by the noise covariance tensor \cite{Pavliotis2014}.}.

We remark that the function $\Phi(\mathbf{x})$ is defined globally but is not, in general, twice differentiable everywhere. Indeed, discontinuities in its first derivatives are present if a) the Hamiltonian associated with the Hamilton-Jacobi equation given in Eq. \ref{eq:HJE} is not integrable (non-integrability being the  typical situation), and b) if the system features more than one co-existing attractors. These latter  discontinuities are of little practical relevance because they appear only for values of $\Phi$ larger than those at the saddles \cite{Graham1986}, for reasons that will become apparent below.

\subsection{Noise-induced Escape from the Attractor}\label{escape}
The quasi-potential $\Phi$ is key for determining the statistics of noise-induced escape from a given attractor. Indeed, the probability that an orbit with initial condition in $B_j$ does not escape from it over a time $p$ decays as:
\begin{equation}\label{eq:tt_distr}
 P_{j,\sigma}(p) =\frac{1}{\tau_{j,\sigma}}\exp\left(-{p}/\tau_{j,\sigma}\right), \quad \tau_{j,\sigma}\sim \exp\left(\frac{2\Delta\Phi_j}{\sigma^2}\right),
\end{equation}
where $\tau_{j,\sigma}$ is the expected escape time and 
%
 $\Delta\Phi_j=\Phi(\Pi_l)-\Phi(\Omega_j)$ is the lowest quasi-potential barrier height \cite{LT:2011}, i.e. $\Phi$ has the lowest value in $\Pi_l$ compared to all the other saddles neighbouring $\Omega_j$. In general, one may need to add a correcting prefactor to $P_{j,\sigma}(p)$ \cite{LT:2011}. 

Note that $\tau_{j,\sigma}$ given in Eq. \ref{eq:tt_distr} does not contain the pre-exponential factor. Ref. \cite{Bouchet2016} provided an expression for such pre-exponential factor for general non-equilibrium diffusion processes under the assumption that attractors and saddles are simple points, thus generalising what is given in \cite{Bovier2004}. As we aim at treating also a more general setting for the geometry of attractors and saddles, we pay below the price of having to take the pre-exponential factors as phenomenological parameters that one can find from experiments or numerical simulations \cite{Bodai2018}. We also remark that, in the zero-noise limit, the transition paths outside a basin of attraction follow the instantons. Instantons are defined as solutions of 
\begin{equation}\label{eqappi}
d{{{x}}_i}/dt={\tilde{F}}_i(\mathbf{x})={R}_i(\mathbf{x})+C_{ij}(\mathbf{x})\partial_j \Phi(\mathbf{x})
\end{equation}
that connect a point in $\Omega_j$ to a point in $\Pi_l$. Instantonic trajectories have a reversed component of the gradient contribution to the vector field compared to regular - relaxation - trajectories.}

Let's now take the simpler case of bistable systems, where we have two attractors $\Omega_1$, $\Omega_2$, and one saddle $\Pi_1$. We can then express the average transitions times as follows: 
\begin{align}\label{eq:tau2}
\tau^{1\rightarrow2}_\sigma \propto \exp\left(\frac{2(\Phi(\Pi_1)-\Phi(\Omega_1)}{{\sigma}^2}\right),\\
\tau^{2\rightarrow1}_\sigma \propto \exp\left(\frac{2(\Phi(\Pi_1)-\Phi(\Omega_2)}{{\sigma}^2}\right),
\end{align}
so that 
\begin{equation}\label{eq:tau3}
\frac{\tau^{1\rightarrow2}_\sigma}{\tau^{2\rightarrow1}_\sigma } \propto \exp\left(\frac{2(\Phi(\Omega_2)-\Phi(\Omega_1)}{{\sigma}^2}\right).
\end{equation}
This implies that, in the weak noise limit, both escape times diverge, but the escape time out of the attractor corresponding to the lower value of the quasi-potential diverges faster. Note that one can expect the proportionality constant in Eq. \ref{eq:tau3} to be $O(1)$. Taking a maximally coarse-grained view on the problem, where we consider the state as represented by the  populations $P_{1,\sigma}$, $P_{2,\sigma}$ of the  neighbourhood of the two attractors, we can write the following master equation:
\begin{align}
\dot{P}_{1,\sigma}&= -\frac{{P}_{1,\sigma}}{\tau^{1\rightarrow 2}_\sigma}+\frac{{P}_{2,\sigma}}{\tau^{2\rightarrow 1}_\sigma}\label{master1}\\
\dot{P}_{2,\sigma}&= -\frac{{P}_{2,\sigma}}{\tau^{2\rightarrow 1}_\sigma}+\frac{{P}_{1,\sigma}}{\tau^{1\rightarrow 2}_\sigma}\label{master2}.
\end{align}
{\color{black}The master equation above makes sense if one assumes the presence of clear timescale separation between the relaxation motions near each attractor and those across the saddle, which depends critically on the presence of weak  noise \cite{Lelievre2015,Gesu2019}.}
 At steady state, we obtain that 
\begin{equation}\label{eq:pop}
\frac{{P}_{1,\sigma}}{{P}_{2,\sigma}}=\frac{\tau^{1\rightarrow2}_\sigma}{\tau^{2\rightarrow1}_\sigma } \propto \exp\left(\frac{2(\Phi(\Omega_2)-\Phi(\Omega_1)}{{\sigma}^2}\right).
\end{equation}
{\color{black}Equation \ref{eq:pop} could be obtained  by  integrating the invariant measure given in Eq. \ref{eq:stationary_distr} }in the neighborouhood of the attractors and  taking a saddle point approximation. Additionally, Eq. \ref{eq:pop} implies that in the weak-noise limit only one of the two deterministic attractors will be populated, and specifically the one where the quasi-potential has lower value. We remark that two different noise laws differing for the correlation matrix $C$ acting on top of the same drift field will define two different quasi-potentials, see Eq. \ref{eq:HJE}. As a result of that, they will in general feature a different selection of the dominating population in the zero noise limit.} One can easily extend the master equation defined above to the case where multiple states and multiple paths of transitions are present. Finally, note that in \cite{Lucarini2019} the mathematical framework described in this section has been used to study stochastic resonance for general non-equilibrium systems.

\section{Numerical Modelling}\label{ModelData}

The climate model considered here is constructed by coupling the primitive equations atmospheric model PUMA \cite{puma} with the Ghil-Sellers energy balance model \cite{Ghil1976}, the latter describing succinctly the meridional oceanic heat transport. It has been already presented in \cite{Lucarini2017} with the name of PUMA-GS, but we report here again its formulation in order to elucidate the role of stochastic forcing, which was absent in the previous version. The stochastic forcing is added as a fluctuating term modulating the value of the incoming radiation determining the energy input into the system. 

\subsection{The Atmospheric Component }

The atmospheric component of the PUMA-GS model is provided by PUMA~\cite{puma}, which consists of a dynamical core: the dry hydrostatic primitive equations on the sphere (mapped laterally by the latitude $\phi$ and longitude $\lambda$), solved by a spectral transform method (only linear terms are evaluated in the spectral domain, nonlinear terms are evaluated in grid-point space). 
The equations for the prognostic state variables, the vertical component (with respect to the local surface) of the absolute vorticity $\zeta=\xi+2\nu\Omega_E$ (where $\xi$ is the vertical component of the relative vorticity, $\nu=\sin\phi$, and $\Omega_E=2\pi$/day is the angular frequency of the Earth rotation) the (horizontal) divergence of the velocity field $D$, the (atmospheric) temperature $T_a=\bar{T_a}+T_a'$ (separated into a time-independent arbitrary reference part $\bar{T_a}$ and anomalies $T_a'$), and the logarithmic pressure (normalized by the surface pressure $p_s$) $\sigma=\ln p/p_s$, read as follows:
\begin{eqnarray}\label{eq:puma}
  \partial_t\zeta &= s^2\partial_{\lambda}F_v - \partial_{\nu}F_u - \tau^{-1}_f\xi - K\nabla^8\xi,\label{eq:puma_zeta} \\
  \partial_t D    &= s^2\partial_{\lambda}F_u + \partial_{\nu}F_v - \nabla^2[s^2(U^2+V^2)/2+\Psi+T_a\ln p_s]\label{eq:puma_D} \\
  &- \tau^{-1}_fD - K\nabla^8D, \nonumber \\
  \partial_t T_a' &= s^2\partial_{\lambda}(UT_a') - \partial_{\nu}(VT_a') + DT_a' - \dot{\sigma}\partial_{\sigma}T_a \label{eq:puma_T} \\
  &+ \kappa T_a\omega/p + \tau_c^{-1}(T_R(T_S) - T_a) - K\nabla^8T_a', \nonumber \\
  \partial_t\ln p_s &= -s^2\partial_{\lambda}\ln p_s - V\partial_{\nu}\ln p_s - D - \partial_{\sigma}\dot{\sigma},\label{eq:puma_p} \\
  \partial_{\ln\sigma}\Psi &= -T_a,\label{eq:puma_state}
\end{eqnarray}
where $s^2=1/(1-\nu^2)$, $F_u = V\zeta - \dot{\sigma}\partial_{\sigma}U - T_a'\partial_{\lambda}\ln p_s$, $F_v = -U\zeta - \dot{\sigma}\partial_{\sigma}V - T_a's^{-2}\partial_{\nu}\ln p_s$, $U=u\cos\phi$, $V=v\cos\phi$, $u$, $v$ being respectively the horizontal and vertical wind velocity components, and $\Psi$ is the geopotential height. 
Equations \ref{eq:puma_zeta},\ref{eq:puma_D}, and \ref{eq:puma_p} express the conservation of momentum, Eq. \ref{eq:puma_T} expresses the conservation of energy, and Eq. \ref{eq:puma_state} is the equation of state.

A number of simple parametrizations are adopted in order to improve the realism and the stability of the model. Firstly, the hyperdiffusion operator $K\nabla^8$  is added to the equations of vorticity, divergence and temperature, to represent {\em subgrid-scale} eddies. Secondly, {\em large-scale} dissipation of vorticity and divergence is facilitated by Rayleigh friction of time scale $\tau_f$. Thirdly, the physics of diabatic heating due to radiative heat transport is parametrized by Newtonian cooling: the temperature field is relaxed (with a time scale $\tau_c$) towards a reference or \textit{restoration} temperature field $T_R$, which can be considered a radiative-convective equilibrium solution. We adopt the following simple expression for the restoration temperature~\cite{puma}:
\begin{eqnarray}\label{eq:res_temp}
 T_R = (T_R)_{tp} + \sqrt{[L(z_{tp} - z(\sigma))/2]^2 + S^2} + L(z_{tp} - z(\sigma))/2,\label{eq:res_temp_1} \\
 (T_R)_{tp} = \langle T_S \rangle - \bar{L}z_{tp}, \label{eq:res_temp_2} \\
 L(\lambda,\phi) = \partial_zT_R=(T_S(\lambda,\phi) - (T_R)_{tp})/z_{tp}, \label{eq:res_temp_3}
\end{eqnarray}
where $(T_R)_{tp}$ and $z_{tp}$ are the temperature and height of the tropopause, respectively, $L$ ($\bar{L}$) is the (average) lapse rate, $\langle T_S \rangle$ is the globally averaged surface temperature, and $z(\sigma)$ is determined by an iterative procedure~\cite{puma}. The above expressions indicate that the restoration temperature profile is \textit{anchored} to the surface temperature $T_S$. However, as Eq. \ref{eq:res_temp_2} indicates, $T_R$ at any one point on the sphere is determined by not only the local (dynamical) surface temperature, but also the global average $\langle T_S \rangle$. We note that $T_a(\sigma=1)$ is obtained by linear extrapolation, according to $T_a(\sigma=1) \approx T_a(\sigma=0.9) + \eta(T_S - T_R(\sigma=0.9))$, $0<\eta<1$. With $\eta=1$ the coupling term 
is $k_3(T_a(\sigma=1)-T_S) \approx k_3(T_a(\sigma=0.9)-T_R(\sigma=0.9))$. Generally $\overline{T_a(\sigma=1)-T_S}\neq 0$ (laterally inhomogeneous heating), but $\overline{\langle T_a(\sigma=1)\rangle}=\overline{\langle T_S \rangle} $, where the overbar denotes averaging with respect to time.

For our setup we choose: $K^{-1} = 0.25$ days, $\tau_c=30$ days, $\tau_f=1$ day, $\bar{L} = 0.0065$ K/m, $z_{tp} = 12000$ m, and $k_3=10^{-4}$. We also adopt a coarse resolution of T21 (i.e., the series of spherical harmonics are triangular-truncated at total wave number 21). This implies the optimal number of Gaussian grid points: $N_{lon}= 2N_{lat}=64$. Finally, we consider $N_{lev}=10$ vertical layers and consider a vanishing orography, so that we have a zonally-symmetric configuration. 
The equations are integrated numerically using a $\Delta t = 1$ [hour] time step size.

\subsection{The Ocean Component and the Stochastic Forcing}\label{oceansec}
The surface temperature  $T_S$ is taken to be governed by the a 2D version of the GS EBM~\cite{Ghil1976,Bodai2014}. This model includes a simple yet effective representation of the ice-albedo feedback, and basically defines the slow manifold of the coupled atmosphere-ocean system. The partial differential equation describing the evolution of the ocean surface temperature field $T_S=T_S(t,\phi,\lambda)$ is:
\begin{equation}
\partial_t T_S(t,\phi,\lambda)= \mu\frac{I(\phi)}{C(\phi)}\frac{S_0^*}{4}(1-\alpha(\phi,T_S))-\frac{O(T_S)}{C(\phi)}-\frac{\bar{D}_\phi[T_S]}{C(\phi)}+\frac{\chi[T_S,T_A]}{C(\phi)}+s.f.,\label{1DEBM}
\end{equation}
where $S_0^*$ is the present solar irradiance\footnote{The factor 4 emerges as a result of the geometry of the Earth-Sun system \cite{saltzman_dynamical}}, $\mu=S^*/S^*_0$ as introduced in Sect. \ref{intro}, while the heat capacity $C(\phi)$ and the geometrical factor $I(\phi)$ are explicitly dependent on $\phi$ only, thus enforcing zonally-symmetric boundary conditions. The albedo $\alpha$  depends  on $\phi$ and, critically, on $T_S$, with a rapid transition from strong albedo for low values of $T_S$ ($\alpha_{max}=0.6$) to weak albedo  for $T_S\gtrsim260$ K ($\alpha_{min}=0.2$), which fuels the positive ice-albedo feedback. Additionally, $O$ is the outgoing radiation per unit area, expressed as a monotonically increasing function of $T_S$ (this is responsible for the negative Boltzmann feedback, taking into account also the greenhouse effect), $\bar{D}_\phi$ is a diffusion operator parametrizing the meridional heat transport, and $\chi$ describes the heat exchange with the atmosphere. See \cite{Bodai2014,Lucarini2017} for further details. 

Finally, the last term on the right hand side $s.f.$ is the stochastic forcing, which is introduced as a {\color{black}random modulation of the solar irradiance given by $\mu S_0^*$. Hence, we have:}
\begin{equation}
s.f.=\sigma s(T_s,\phi,\lambda)\frac{dW}{dt}= \sigma \mu\frac{I(\phi)}{C(\phi)}\frac{S_0^*}{4}(1-\alpha(\phi,T_S))\frac{dW}{dt},\label{1DEBMnoise}
\end{equation}
where $\sigma$ controls the intensity of the noise, $s$ defines the noise law, and $dW$ is the increment of a one-dimensional Wiener process. 
 %
%
Since $s$ depends explicitly on $T_S$ via the term $\alpha(\phi,T_S)$,  we are dealing with a multiplicative noise law. We consider the It\^o convention for noise, {\color{black}so that our (discretised) equations are in the form of Eq. \ref{eqapp}. See a discussion on the relevance of the chosen convention in Sect. \ref{itovsstrato}.} Adding a Gaussian random variable of variance $\sigma$ at each time step $\Delta t$ ($1$ hour ) of the model amounts to considering that, on the time scale $\tau=N\times \Delta t$, the relative fluctuation of the solar irradiance scales as $\sigma_\tau=\sigma/\sqrt{N}$. 

%
{\color{black}
As mentioned in Sect. \ref{stochastic}, our approach  requires assuming the validity of the hypoellipticity condition. In order to prove this, we should test the H\"ormander condition for the evolution equations of the model. This is of great relevance but is beyond the specific scope of this paper, while indeed deserving a separate and accurate investigation. However, as discussed below we can heuristically understand why, indeed, it is reasonable to assume that stochastic forcing acting on the oceanic surface temperature 
propagates to all degrees of freedom of the coupled system, as usually implicitly assumed in basically any numerical study of stochastically forced geophysical flows \footnote{Obviously,  numerical truncation introduces some additional noise on all degrees of freedom of the system.}. 

We can first approach this problem by looking at the structure of the evolution equations.  
%
%
%
%
%
%
The stochastic forcing given in Eq. \ref{1DEBMnoise} impacts directly the $T_S$ field, as shown in Eq. \ref{1DEBM}. The $T_S$ fields determines the restoration temperature $T_R$, see Eq. \ref{eq:res_temp}. In turn, the restoration temperature impacts the anomaly of the atmospheric temperature field $T_a'$ (see Eq. \ref{eq:puma_T}), which in turn affects the vorticity field $\zeta$ - Eq. \ref{eq:puma_zeta} - and divergence field $D$ - Eq. \ref{eq:puma_D}. Finally, anomalies in $D$ impact the surface pressure $p_s$, as clear from Eq. \ref{eq:puma_p}. The nonlinear terms corresponding to advective processes on the right hand side of each Eq. \ref{eq:puma_zeta}-\ref{eq:puma_p}, which contain two more of the above mentioned fields, make sure that noise propagates across all scales in each field. This latter point could be better understood by taking a truncated Fourier representation of Eqs. \ref{eq:puma_zeta}-\ref{eq:puma_p}, which, in fact, closely corresponds to the actual formulation of the numerical model implemented here. Concluding, no dynamical or thermodynamical field and no scale within each field is insulated from the noise, even if the covariance matrix of the noise law is \textit{extremely} singular, as noise impacts directly only a small fraction of the degrees of freedom of the system.

On more physical grounds, one can observe that the ocean surface temperature drives the restoration profile of the atmospheric temperature, and that fluctuations of such a profile modulate the atmospheric instabilities, whose energy cascades down to the smallest scales resolved by the model; see \cite{Peixoto1992,Lucarini2014b,Ghil2019} for a detailed treatment of the energetics of the climate system.


}

\section{Results}\label{Results}
\begin{figure}
a)\includegraphics[trim=0.5cm 0.5cm 0.5cm 0.5cm,  clip=true, width=0.75\textwidth]{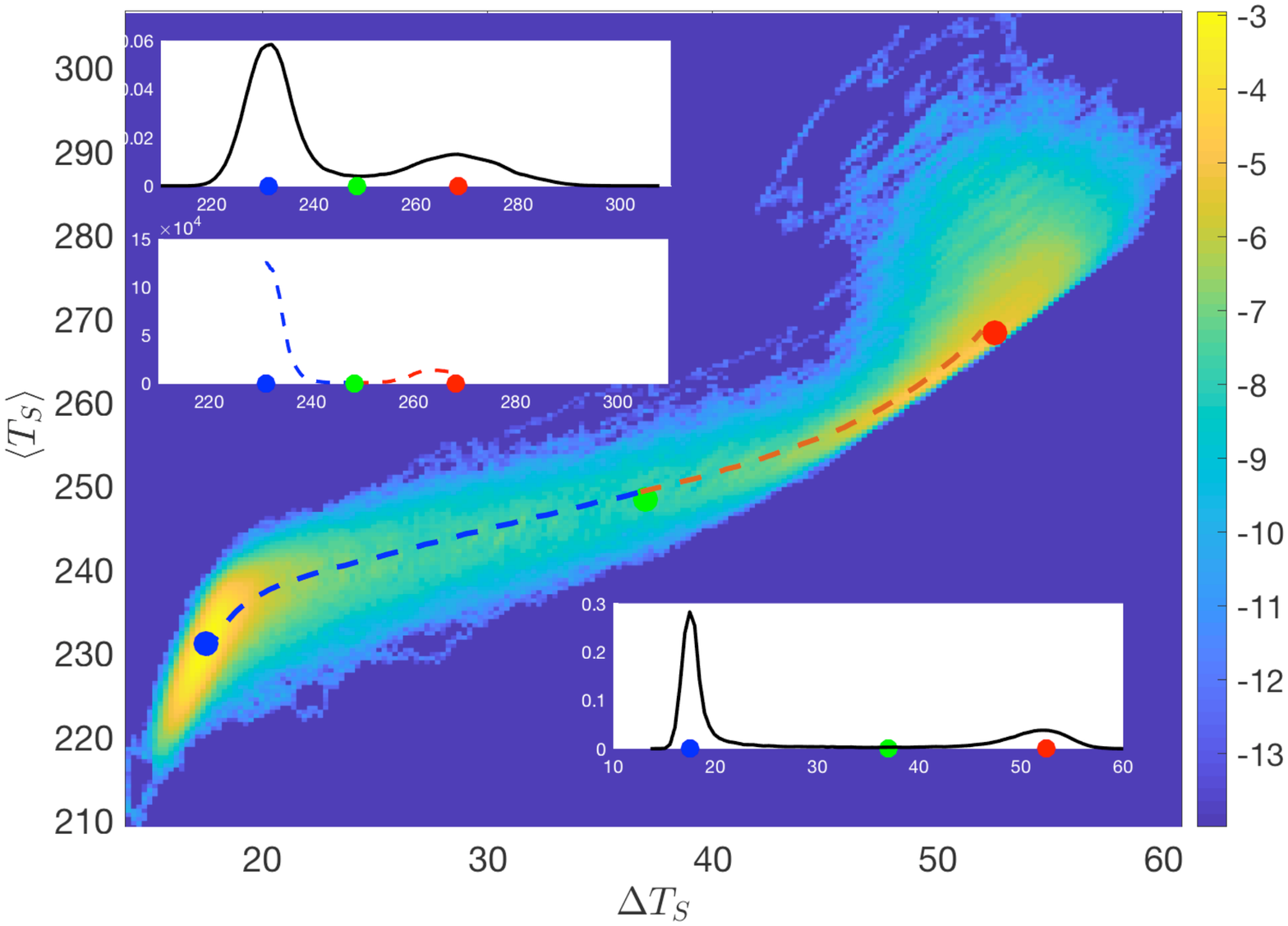}\\
b)\includegraphics[trim=0.5cm 0.5cm 0.5cm 0.5cm,  clip=true, width=0.75\textwidth]{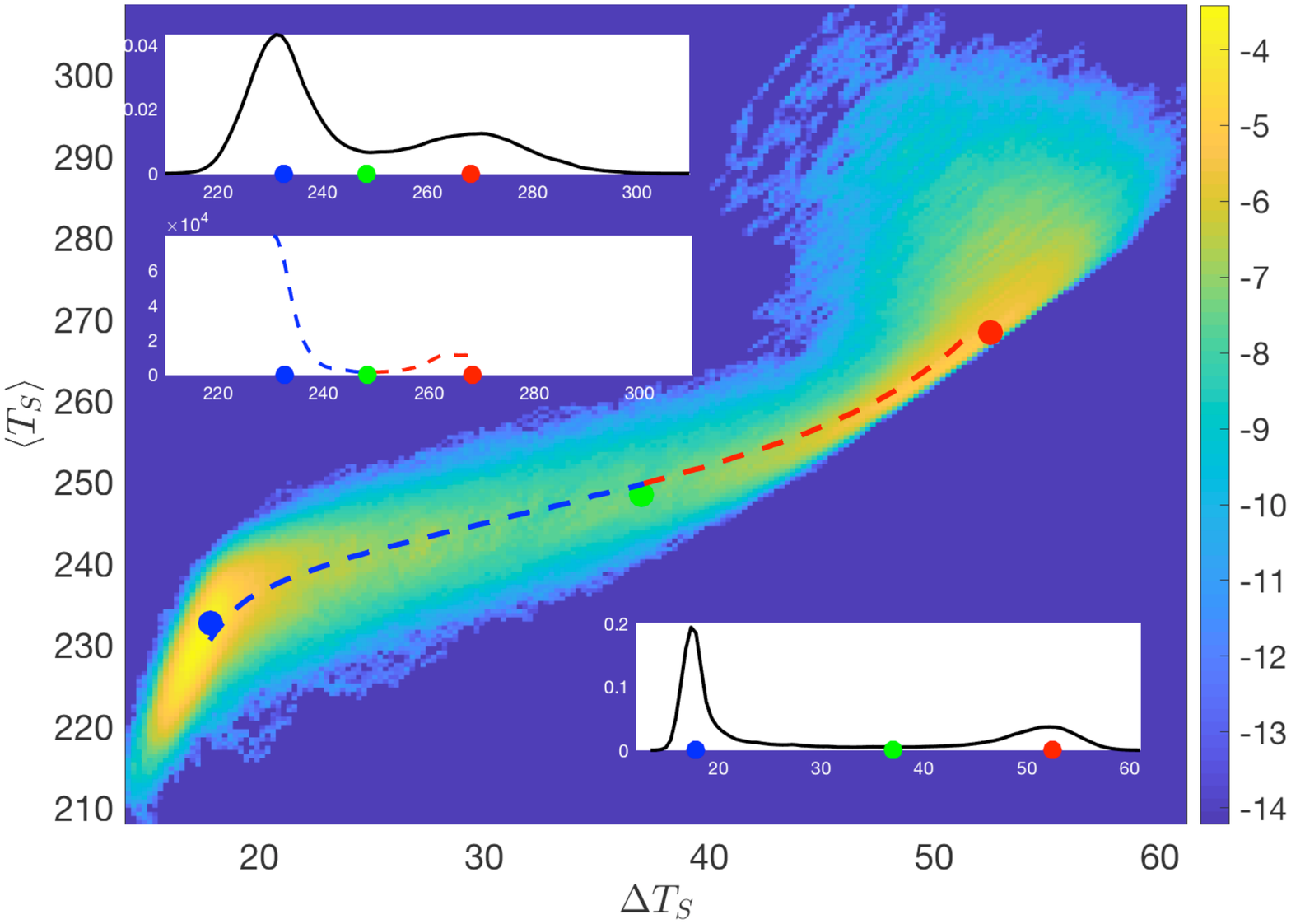}%
\caption{Panel a) Main graph: density in the projected phase space $(\Delta T_S,\langle T_S \rangle)$ {\color{black}(in units of $K$ in both axes)}, with indication of the actual position of the W attractor (red dot), SB attractor (blue dot), M state (green dot) for $\mu=1$. The  $W\rightarrow SB$ and $SB\rightarrow W$ approximate instantons are also indicated. We  have used $\sigma_{100y}=1.5\%$.  Top left inset: marginal pdf with respect to $\langle  T_S \rangle$. Bottom right inset: marginal pdf with respect to $\Delta T_S $.  Center left inset: pdf along the path of the two instantons. Reproduced from \cite{LB2018}. Panel b): same as panel a), with  $\sigma_{100y}=1.8\%$. \label{mu100_invariant}}
\end{figure}

We first treat in detail three  cases inside the region of bistability depicted in Fig. \ref{oldbifurcation}, namely $\mu=0.98$ (close to the tipping point $\mu_{W\rightarrow SB}$),  $\mu=1.0$ (corresponding to present-day solar irradiance), and $\mu=1.02$ (in the parametric region where the M state undergoes a symmetry-break bifurcation). We then construct the weak-noise limit of the invariant measures for all the values of $\mu$ in the region of bistability. We remark that the results shown in Figs. \ref{mu098tau}, \ref{mu098}, and \ref{mu100_invariant}a) have already been reported in the short communication \cite{LB2018}, but we deem extremely useful to present them here as well because they are now part of a much more complex, coherent, and detailed narrative. 

\subsection{Escapes from Basins of Attraction and Instantons}
\begin{figure}
\includegraphics[trim=0cm 0.5cm 0cm 0.5cm, clip=true, width=0.7\textwidth]{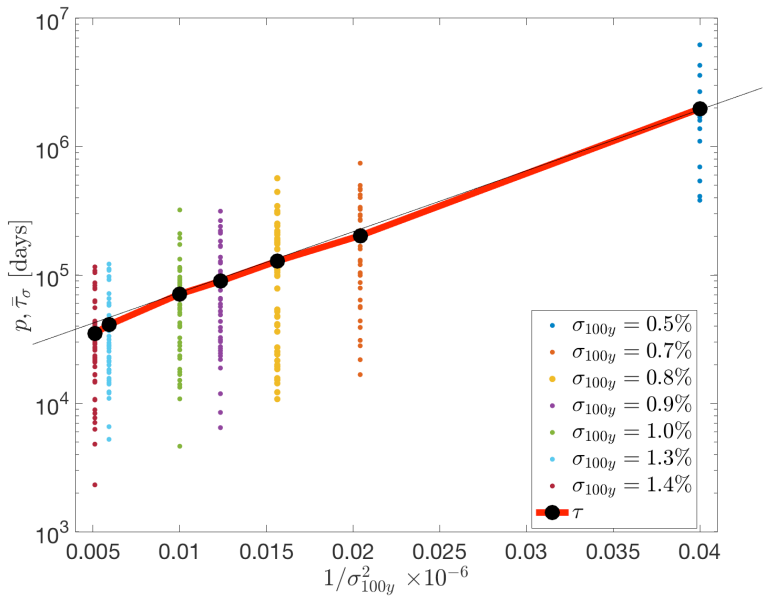}%
\caption{Statistics of the escape times $p$ for the noise-induced $W\rightarrow SB$ transitions for various noise strengths. Each coloured dot corresponds to an observed escape time $p$. The estimate of the expected escape times $\bar{\tau}_\sigma$ are indicated by the black dots connected by the red line. The slope of the straight line fit gives the potential difference described in Eq. \ref{eq:tau2}. See \cite{Bodai2018} for an optimal algorithm for estimating the potential difference.  Reproduced from \cite{LB2018}. \label{mu098tau}}
\end{figure}

\begin{figure}
\includegraphics[trim=0cm 0.0cm 0.cm 0.cm, clip=true, width=0.7\textwidth]{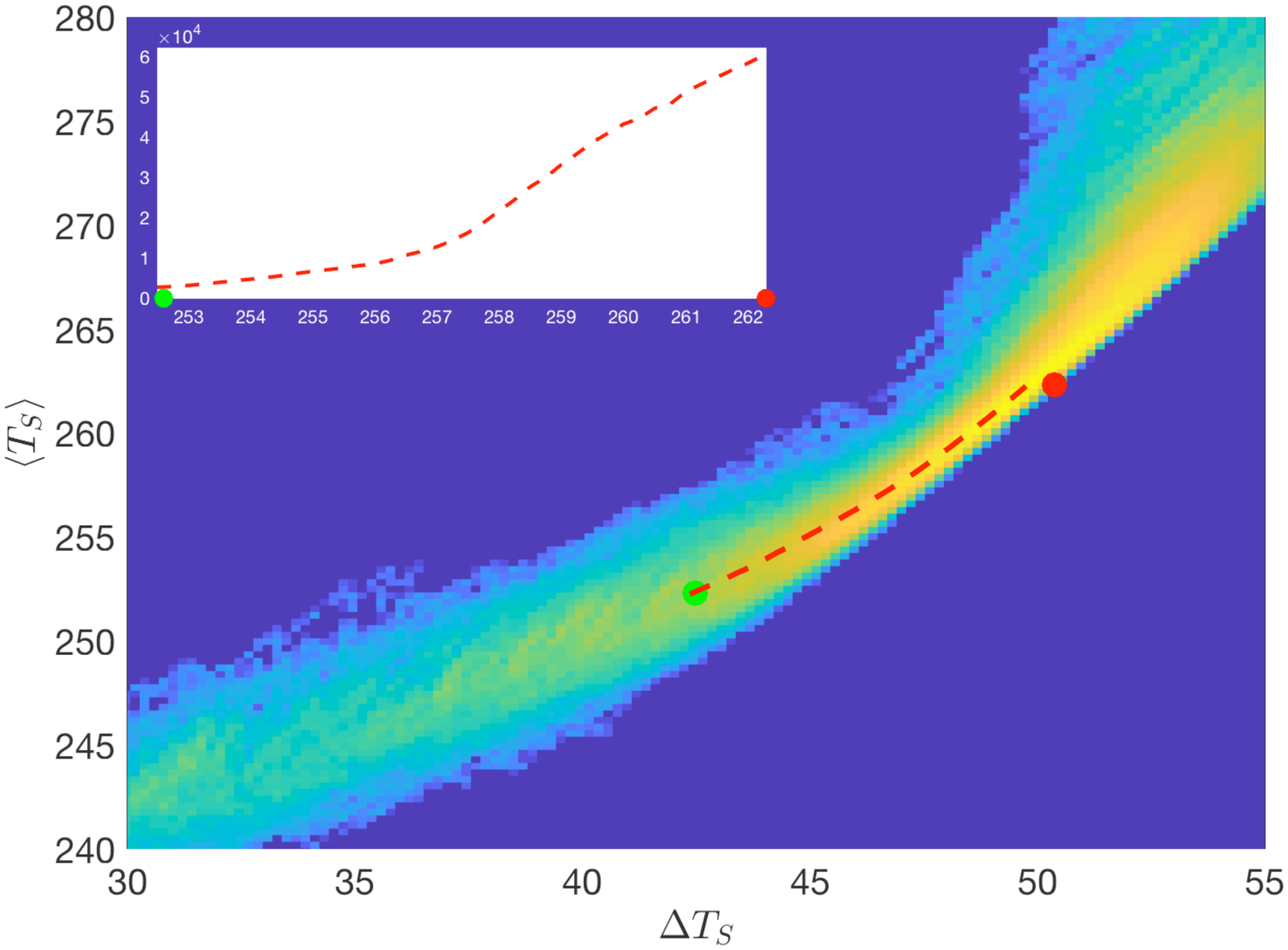}%
\caption{Main graph: Logarithm of the transient density $\tilde\rho$ in the reduced space $(\Delta T_S,\langle T_S \rangle)$ {\color{black}(in units of $K$ in both axes)}, with indication of the actual position of the W attractor (red dot) and  the M state (green dot) for $\mu=0.98$. We  have used $\sigma_{100y}=1\%$. The  $W\rightarrow SB$ approximate instanton is indicated. Top left inset: pdf along the path of the instanton ($\langle T_S \rangle$ on the x-axis). Reproduced from \cite{LB2018} \label{mu098}.}
\end{figure}

\begin{figure}
\includegraphics[trim=0cm 0cm 0cm 0cm, clip=true, width=0.7\textwidth]{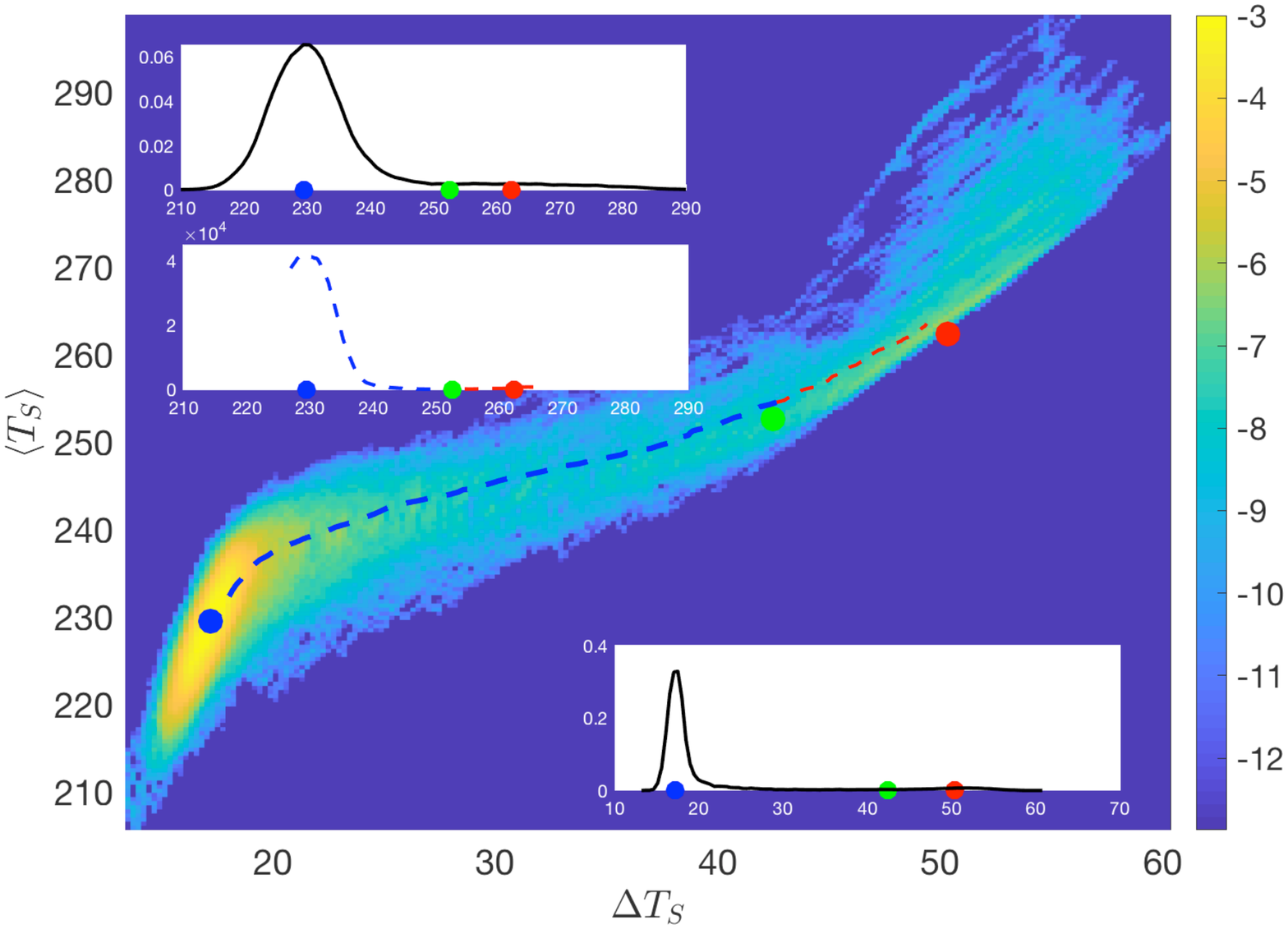}%
\caption{Main graph: density in the projected phase space $(\Delta T_S,\langle T_S \rangle)$  {\color{black}(in units of $K$ in both axes)}, with indication of the actual position of the W attractor (red dot), SB attractor (blue dot), M state (green dot) for $\mu=0.98$. The  $W\rightarrow SB$ and $SB\rightarrow W$ approximate instantons are also indicated. We  have used $\sigma_{100y}=1.5\%$.  Top left inset: marginal pdf with respect to $\langle  T_S \rangle$. Bottom right inset: marginal pdf with respect to $\Delta T_S $.  Center left inset: pdf along the path of the two instantons. \label{mu098_invariant}}
\end{figure}
In the case of $\mu=0.98$, we first perform a set of simulations with noise of different intensity rangiW ng from $\sigma_\tau=0.5\%$ to $\sigma_\tau=1.4\%$, with $\tau=100$ years ($y$). For each value of the noise intensity, we initialise 50 trajectories in the basin of attraction of the W climate and study the statistics of the escape time to the SB attractor. When the transition takes place, we stop the integrations. We observe (not shown) that for each value of $\sigma_\tau$ the escape times are to a good approximation exponentially distributed, thus obeying Eq. \ref{eq:tt_distr}; the process of transition behaves like a Poisson process. The results on the expectation value of the transition times are presented in Fig. \ref{mu098tau}, where we show that, indeed, $\tau_\sigma$ agrees with the prediction of Eq. \ref{eq:tt_distr}. Hence, it is possible to define the difference between the value of the quasi-potential $\Phi$ at the M state and at the W attractor as the slope of the straight line. For reference, we have that  for $\sigma_{100 y}=0.5\%$ the average escape time is about $5.2\times10^3$ $y$. We can predict that the escape rate increases to about $1.5\times10^7$ $y$ when $\sigma_{100 y}\sim0.3\%$. We have that the slope of the straight line in Fig. \ref{mu098tau} gives $\sim2(\Phi(M)-\Phi(W))$. Therefore, we show that it is indeed possible to estimate quantitatively the properties of the quasi-potential also in a very high-dimensional dynamical system like the one considered here. The operation can be repeated for all the other values of $\mu$ in the range of multistability and for the processes of escape from the SB attractor, but we do not pursue here a systematic study of this. 

We then wish to look at the paths corresponding to the transitions. Following the discussion in \cite{Bodai2014,Lucarini2017}, we choose to consider the reduced phase space spanned by the globally averaged surface ocean temperature $\langle T_S\rangle $ and by the meridional temperature difference $\Delta T_S$, defined as the difference between the spatially averaged ocean temperature field between the Equator and $30 ^\circ N$ and between  $30 ^\circ N$ and the North Pole. This reduced phase space provides a minimal yet physically informative viewpoint on the problem, because it is directly linked with the  main physical processes occurring in the climate model:
\begin{itemize}
\item The average surface temperature $\langle T_S\rangle $ is directly associated to the positive ice-albedo feedback and the negative Bolzmann radiative feedback;
\item The meridional temperature difference $\Delta T$ controls the meridional heat transport performed by the ocean, as a result of the diffusive law we insert into its evolution equation;
\item The meridional temperature gradient $\Delta T$ also controls the meridional heat transport performed by the atmosphere, as a result of the mechanism of baroclinic instability \cite{Holton}.
\end{itemize}
Figure \ref{mu098} depicts, for the case $\sigma_\tau=1.0\%$, the transient two-dimensional distribution function $\tilde\rho$ constructed using a frequentist approach using the 50 simulations described above, where the  statistics is collected only until the $W\rightarrow SB$ transition is realised. The distribution we obtain cannot be interpreted as an approximation of the invariant measure, because the integrations are stopped after the transitions. Nonetheless, it is apparent that  the transitions take prominently place in a very narrow band linking the W attractor  and the M state\footnote{Note that, even if the W attractor and the M state look like dots, they have, in fact, a finite (yet very small) size, because they are both chaotic (see caption of Fig. \ref{oldbifurcation}). Here we are considering oceanic variables, which feature a very small variability in the deterministic chaotic case.}. 
In order to obtain a better understanding of the transition paths, we construct an estimate of the instanton linking the W attractor to the M state and associated to the $W\rightarrow SB$ transitions by conditionally averaging the trajectories according to the value of $\langle T_S\rangle$. To a good approximation, the instanton connects the W attractor to the M state, and follows a path of decreasing  density. We do not find evidence of different paths for the trajectories leading to an escape and the relaxation trajectories, which is, instead, a typical signature of non-equilibrium \cite{ZinnJustin1996}. This can be explained by  considering \cite{Bodai2014}, where it is shown that the ocean model evolve to a good approximation in an energy landscape. See also Section \ref{appa}.

\begin{figure}
\includegraphics[trim=0.5cm 0.5cm 0.5cm 0.5cm, clip=true, width=0.75\textwidth]{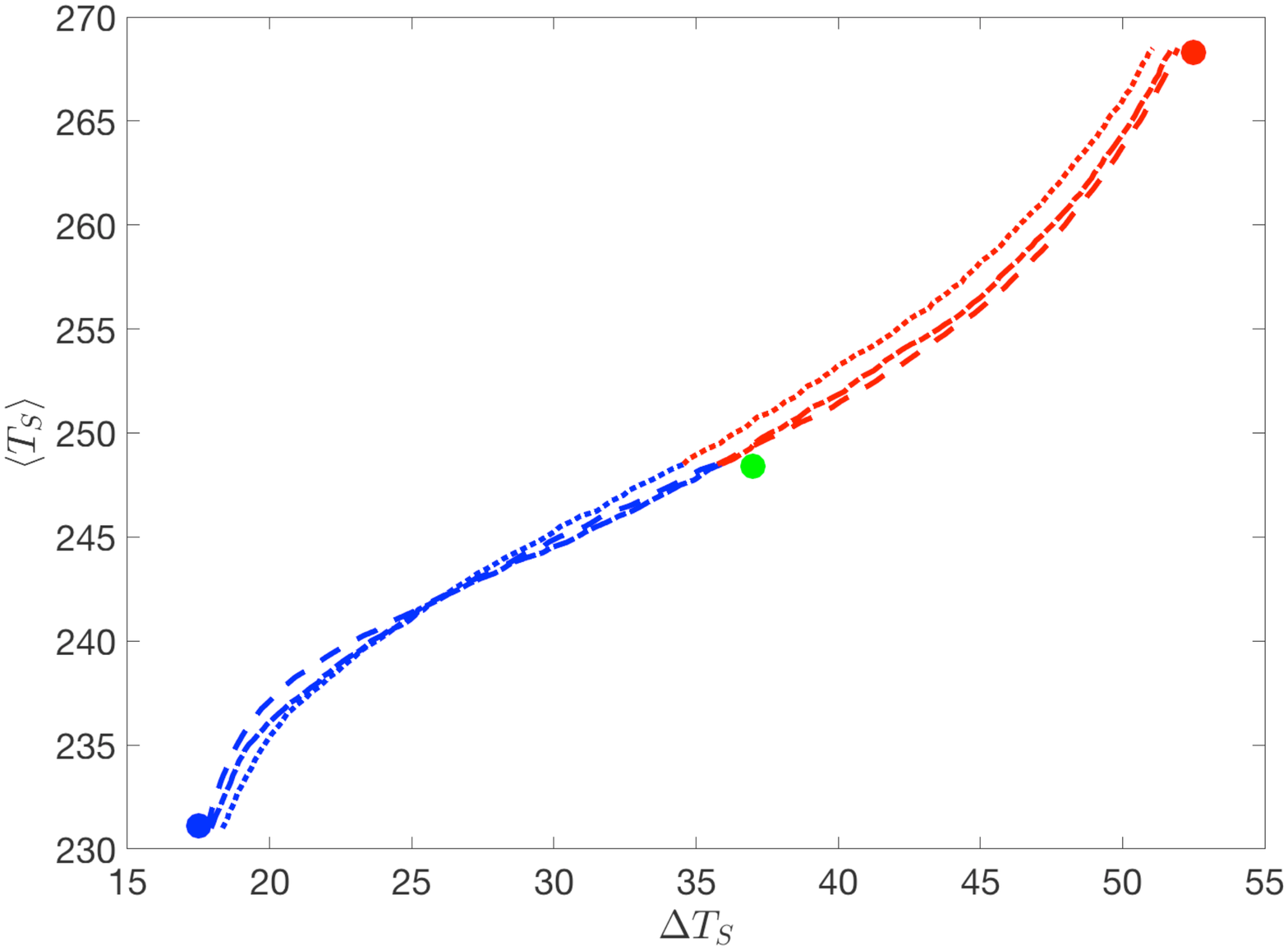}%
\caption{Estimate of the instantons for $\mu=1$ obtained using $\sigma_{100y}=1.5\%$ (dashed lines), $\sigma_{100y}=1.8\%$ (dash-dotted lines), and $\sigma_{100y}=2.5\%$ (dotted lines) {\color{black}(units of $K$ in both axes)}. The red (blue) lines show the estimates for the $W\rightarrow SB$ ($SB\rightarrow W$) instanton.  The dots indicate the actual position of the W attractor (red dot), SB attractor (blue dot), and M state (green dot). The estimate of the instanton improves as the intensity of the noise is reduced.}\label{instantonsmu100}
\end{figure}

\begin{figure}
\includegraphics[trim=0.5cm 0.5cm 0.5cm 0.5cm, clip=true, width=0.75\textwidth]{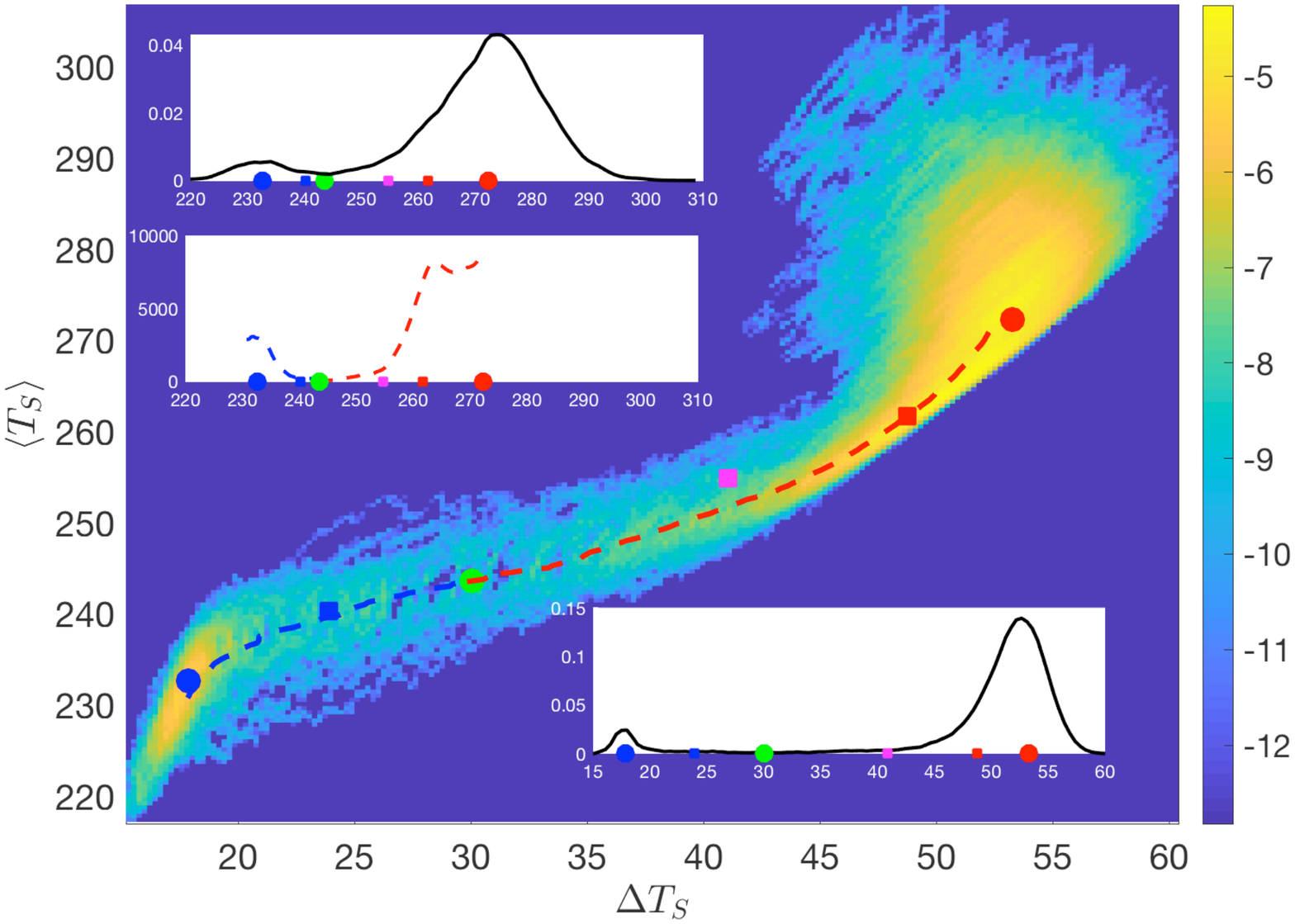}%
\caption{Main graph: density in the projected phase space $(\Delta T_S,\langle T_S \rangle)$, with indication of the position of the W attractor (red dot), SB attractor (blue dot), transient M state (green dot) for $\mu=1.02$ {\color{black}(units of $K$ in both axes)}. The squares indicate the properties of the asymptotic M state: W sector (red square); cold sector (blue square); average (magenta square). The  estimates of the $W\rightarrow SB$ and $SB\rightarrow W$  instantons are also indicated as the red and the blue line, respectively. We  have used $\sigma_{100y}=1.5\%$.  Top left inset: marginal pdf with respect to $\langle  T_S \rangle$. Bottom right inset: marginal pdf with respect to $\Delta T_S $.  Center left inset: pdf along the path of the two instantons.  \label{mu102_invariant}}
\end{figure}

\subsection{Construction of the Invariant Measure}
The information contained in Fig. \ref{mu098} is limited because we are studying only $W\rightarrow SB$ escape processes, and we do not allow for the establishment of an invariant measure. The problem lies in the fact that the quasi-potential minimum associated to the $SB$ attractor is much deeper than the one associated to the $W$ attractor, so that the average escape time associated to $SB\rightarrow W$ transitions is prohibitively long for the range of (rather weak) noise intensities used for constructing Fig. \ref{mu098}. In fact, in order to be able to construct the invariant measure of the system, we need to observe a sufficient number of $W\rightarrow SB$ and $SB\rightarrow W$ transitions, in order to be sure that we have collected a satisfactory statistics; see also the master equations for the populations given in Eqs. \ref{master1}-\ref{master2}. We next increase the noise intensity by setting $\sigma_{100y}=1.8\%$, so that, in an integration lasting about $1.0\times10^4$ $y$, we observe 10 $SB\rightarrow W$ and $W\rightarrow SB$ transitions.  The number of transitions is low because it is extremely hard to escape from the $SB$ state. 

Our results are shown in Fig. \ref{mu098_invariant}. We portray the logarithm of the projection of the invariant measure on the $(\Delta T_S,\langle T_S \rangle)$ plane; we refer to this also as the two-dimensional probability distribution function (pdf). We have that the peaks of the pdf are in good agreement with the position of the $W$ and $SB$ attractors, as implied by the large deviation result presented in Eq. \ref{eq:stationary_distr}. The agreement is even clearer when considering the two marginal pdfs, constructed by projecting  the invariant measure on the one dimensional spaces defined by  $\Delta T_S$ and  $\langle T_S \rangle$. Note that the peak over the $W$ state is hardly noticeable because the occupation of the state is extremely low (less than 5\% of the total); compare with the cases studied below where higher values of $\mu$ are considered. We are also able  to construct both the $W\rightarrow SB$ and the $SB\rightarrow W$ instantons, whose starting points agree remarkably well with the position of the $W$ and the $SB$ attractors, respectively, while their final points are in good agreement with the position of the  M state. By constructing the pdf along the instantons, we find that they  follow a path of monotonic descent (indeed, they follow closely the crests of the pdf), with the minimum located at the M state. Again, this last property can be hardly visualised in Fig. \ref{mu098} for the  $W\rightarrow SB$ instanton (see inset), because the population near the $W$ state is quite small.


Next, we repeat the analysis for $\mu=1$; results are shown in Figs. \ref{mu100_invariant}a)-b). In Panel a) we consider an integration with $\sigma_{100y}=1.5\%$ lasting about $2.9\times10^4$ $y$ and characterised by 41 $SB\rightarrow W$ and $W\rightarrow SB$ transitions. We have an occupation of about $30\%$ for the $W$ basin of attraction, and of about $70\%$ for the $SB$ basin of attraction; additionally, we have $\tau_\sigma^{W\rightarrow SB}\sim 210 y$ and  $\tau_\sigma^{SB\rightarrow W}\sim 460 y$. Also in this case, the projection  of the invariant measure in the $(\Delta T_S, \langle T_S\rangle)$ plane shows that there is good agreement between the position of the peaks of the pdfs and the attractors, and that the estimates of the instantons connect attractors and M states with a good precision. It is also clear that the instantons follow a path of descent in terms of probability, as shown by the central inset.

In Panel b) we show the results of repeating the analysis for $\sigma_{100y}=1.8\%$. In this case the simulation lasts about $2.7\times10^4$ $y$ and we obtain 73 $SB\rightarrow W$ and $W\rightarrow SB$ transitions, and we can draw similar conclusions as in Panel a) regarding the relative position of the attractors, of the M state, and of the instantons. The marginal pdfs are clearly less peaked than in Panel a); the occupancy rate changes slightly with respect to the previous case: it is about  $35\%$ for the $W$ basin of attraction, and of about $65\%$ for the $SB$. 
Instead, the average escape times change more substantially, and can be estimated as $\tau_\sigma^{W\rightarrow SB}\sim 160 y$ and  $\tau_\sigma^{SB\rightarrow W}\sim 300 y$. These last two results indicate that the difference between the value of the quasi-potential at the two competing attractors is relatively small. We will explore this matter in Sect. \ref{measure}, where we will try to deduce where the quasi-potential  reaches its absolute minimum for each value of $\mu$ in the bistable region.

It is worth looking more in detail at how the estimate of the instantons  is impacted by the intensity of the noise used in the simulation. As instantons are defined in the weak-noise limit, we would expect that one achieves higher precision when weaker noise is used. This is confirmed by the results shown in Fig. \ref{instantonsmu100}: we have that the estimates of the instantons obtained using lower noise intensity come closer to the attractors and to the M state. Nonetheless, also the instantons obtained for very strong noise are still relatively accurate. 

\subsection{Instantons and Transitions across the Symmetry-broken Melancholia State}

We next examine the noise-induced transitions for $\mu=1.02$. This case is quite interesting because, as discussed in \cite{Lucarini2017} and reported in Fig. \ref{oldbifurcation}, the longitudinally-symmetric M state  is transient with a very long life time, and slowly evolves into a symmetry broken M state featuring a relatively cold and a relatively warm region, separated by two small regions with large longitudinal temperature gradient at all latitudes. It seems relevant to test whether noise-induced transitions take place through the transient, symmetric M state or the true, symmetry-broken one. Results are shown in Fig. \ref{mu102_invariant}, where we use data from a simulation lasting about $10^4$ $y$ with $\sigma_{100y}=1.5\%$. We observe only 6 transitions in both directions. This time, as opposed to the case of $\mu=0.98$, the figure is so low because it is extremely hard to escape from the $W$ state. We estimate the escape times as $\tau_\sigma^{W\rightarrow SB}\sim 1400 y$ and  $\tau_\sigma^{SB\rightarrow W}\sim 140 y$. We portray the logarithm of the invariant measure of the system in the usual projected space, and the estimates of the instantons. We first observe that in this case most of the density is concentrated around the W attractor, and a nontrivial relation exists between the paths of the instantons and the dynamical structures on the basin boundary. It is clear that the transient M state plays the role of the gateway of transitions as seen in the previous cases, despite  its transient nature. The noise-induced transitions do not go through the actual M state (magenta square), while they seem to go thorough states resembling the properties of the W (red square) and cold (blue square) sectors in the asymptotic M state. This feature might result from the consideration of noise with finite (and not infinitesimal) strength: the quasi-potential near the transient M state might be just barely higher than that of the asymptotic M state (and possibly with a more favourable pre-exponential factor), so that a small but finite noise perturbation might push an orbit near the transient M state into the other basin of attraction. {\color{black}In order to address this point and find instantonic paths connecting the attractors with the true M states, one might need to resort to using more sophisticated numerical techniques. In particular, one should consider using rare events algorithms \cite{rubino2009rare,Ragone2017} to rigorously construct instantonic trajectories \cite{Grafke_2013}. This is beyond the current abilities of the authors but definitely deserves attention in future studies.}

\subsection{Selection of the Limit Measure in the Weak-noise Limit and First-order Phase Transition}\label{measure}
\begin{figure}
a)\includegraphics[trim=0.5cm 0.5cm 0.5cm 0.5cm, clip=true, width=0.47\textwidth]{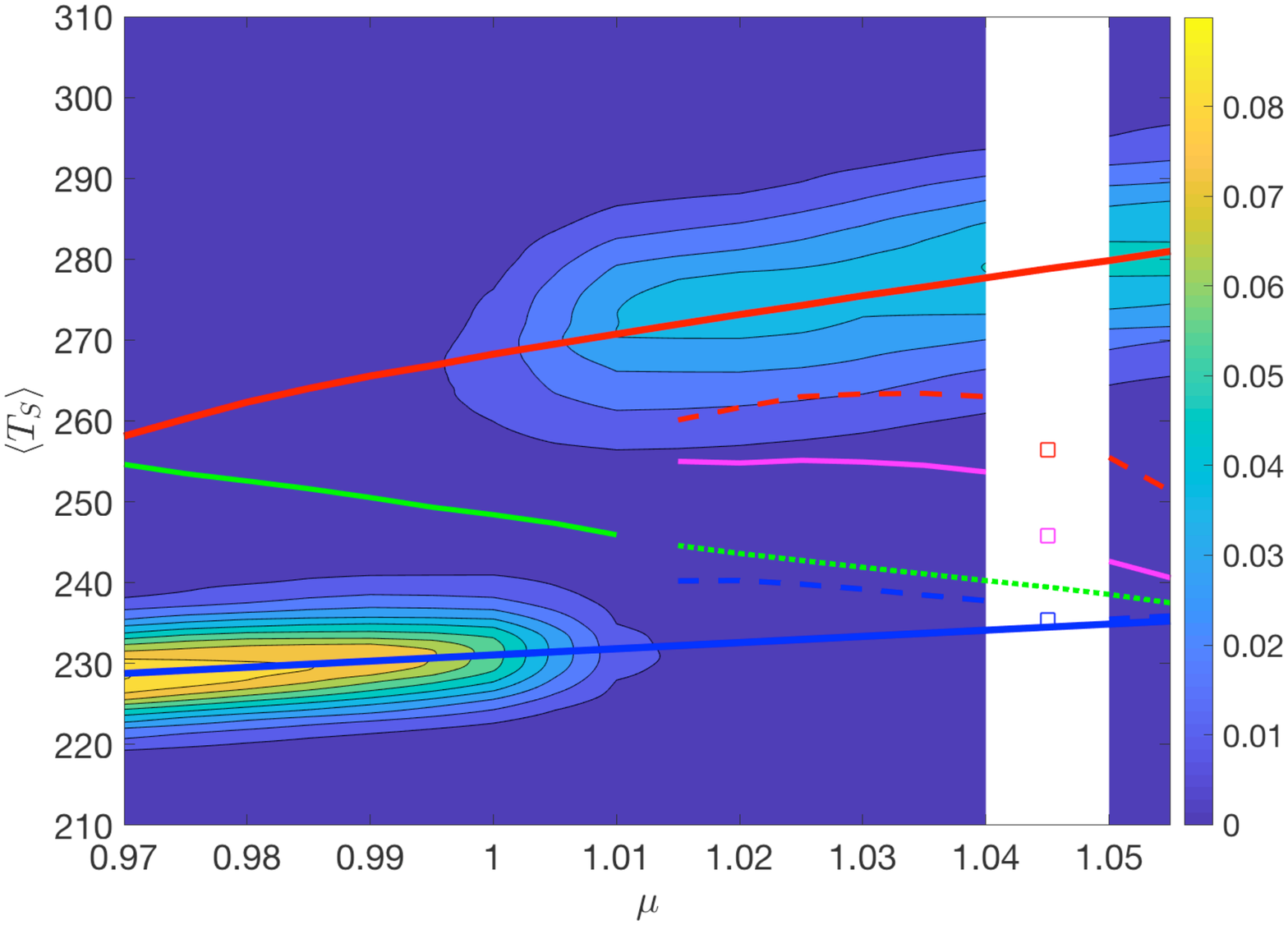}%
b)\includegraphics[trim=0.5cm 0.5cm 0.5cm 0.5cm, clip=true, width=0.47\textwidth]{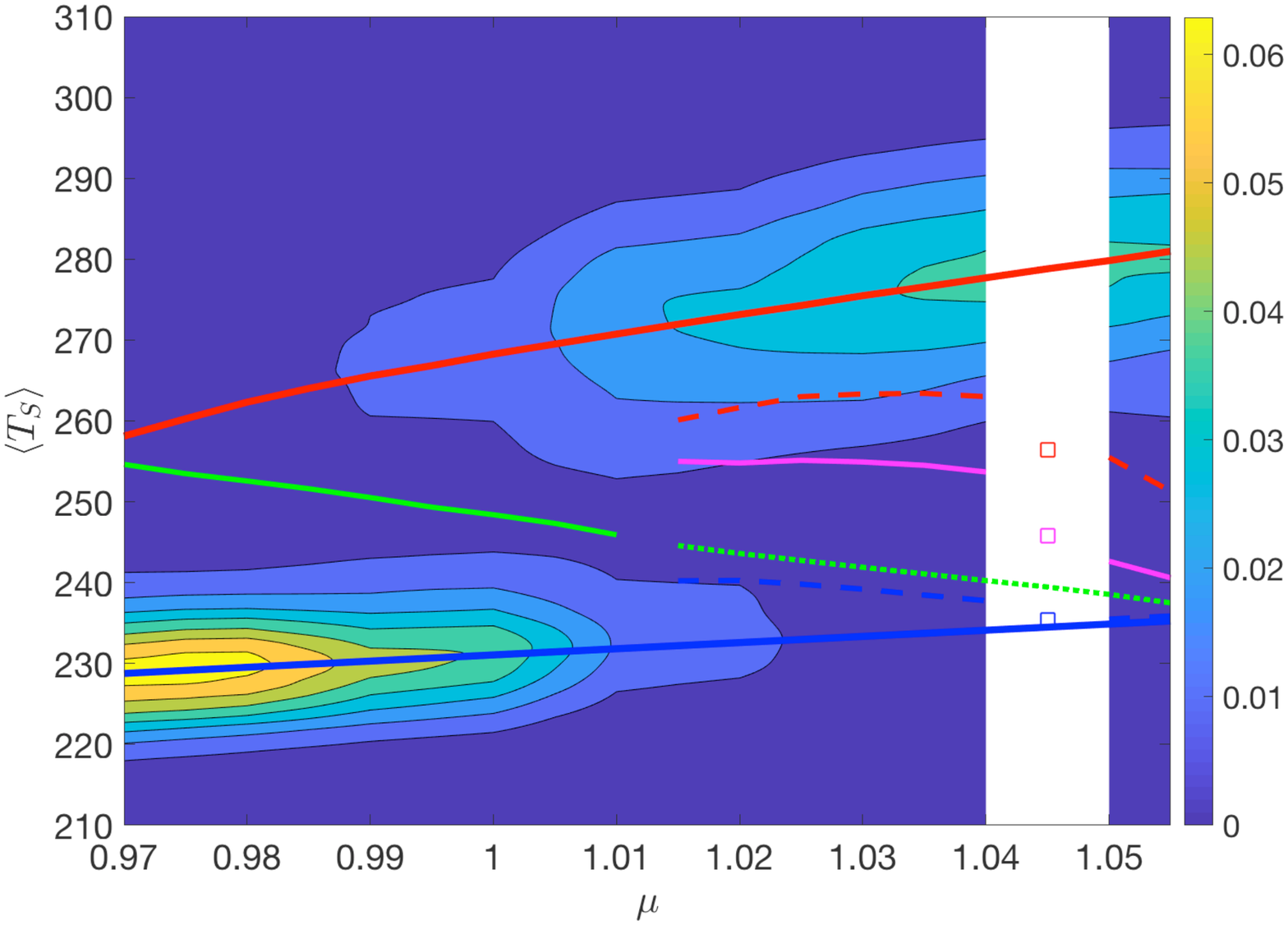}\\
c)\includegraphics[trim=0.5cm 0.5cm 0.5cm 0.5cm, clip=true, width=0.47\textwidth]{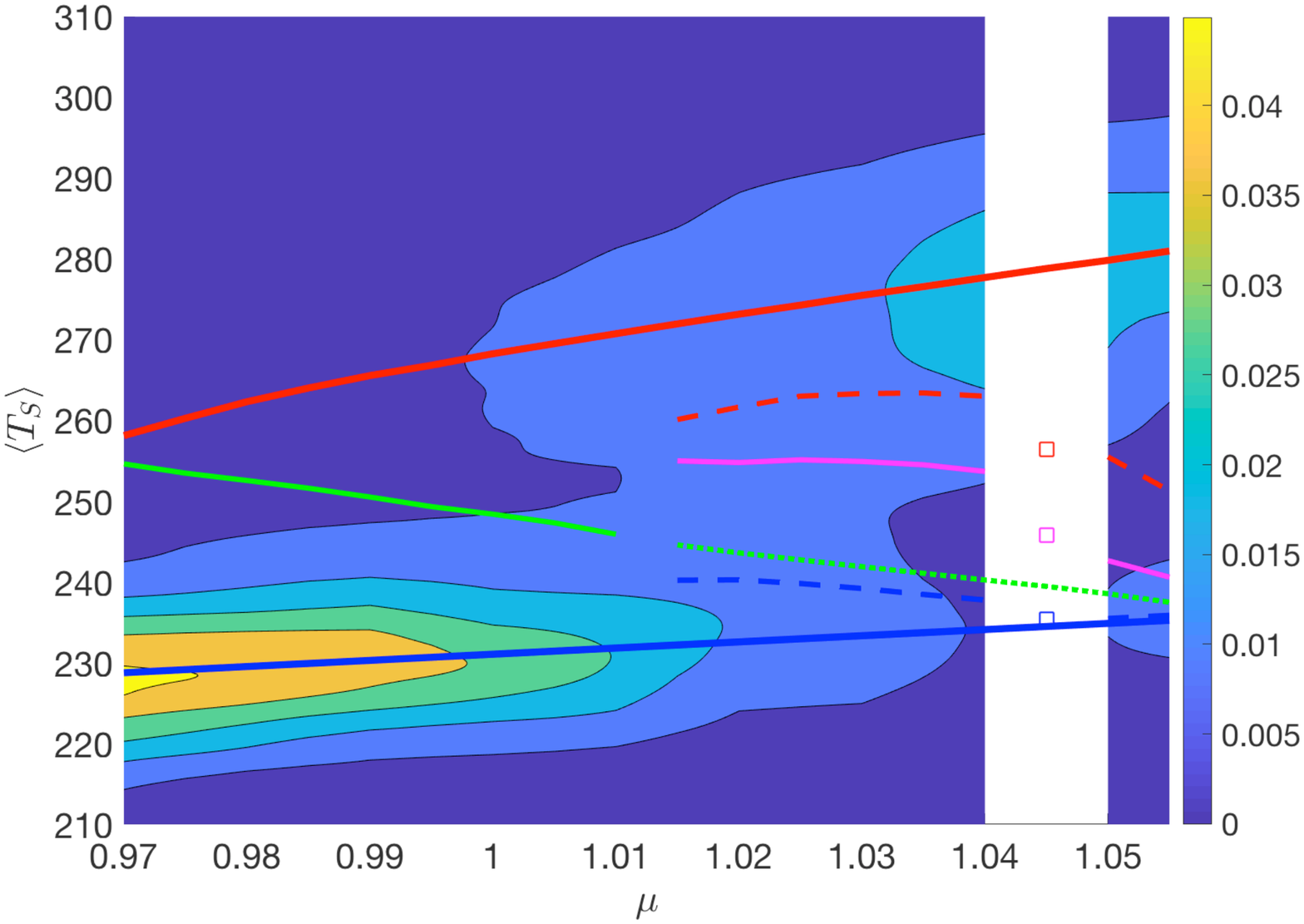}
d)\includegraphics[trim=0.5cm 0.5cm 0.5cm 0.5cm, clip=true, width=0.47\textwidth]{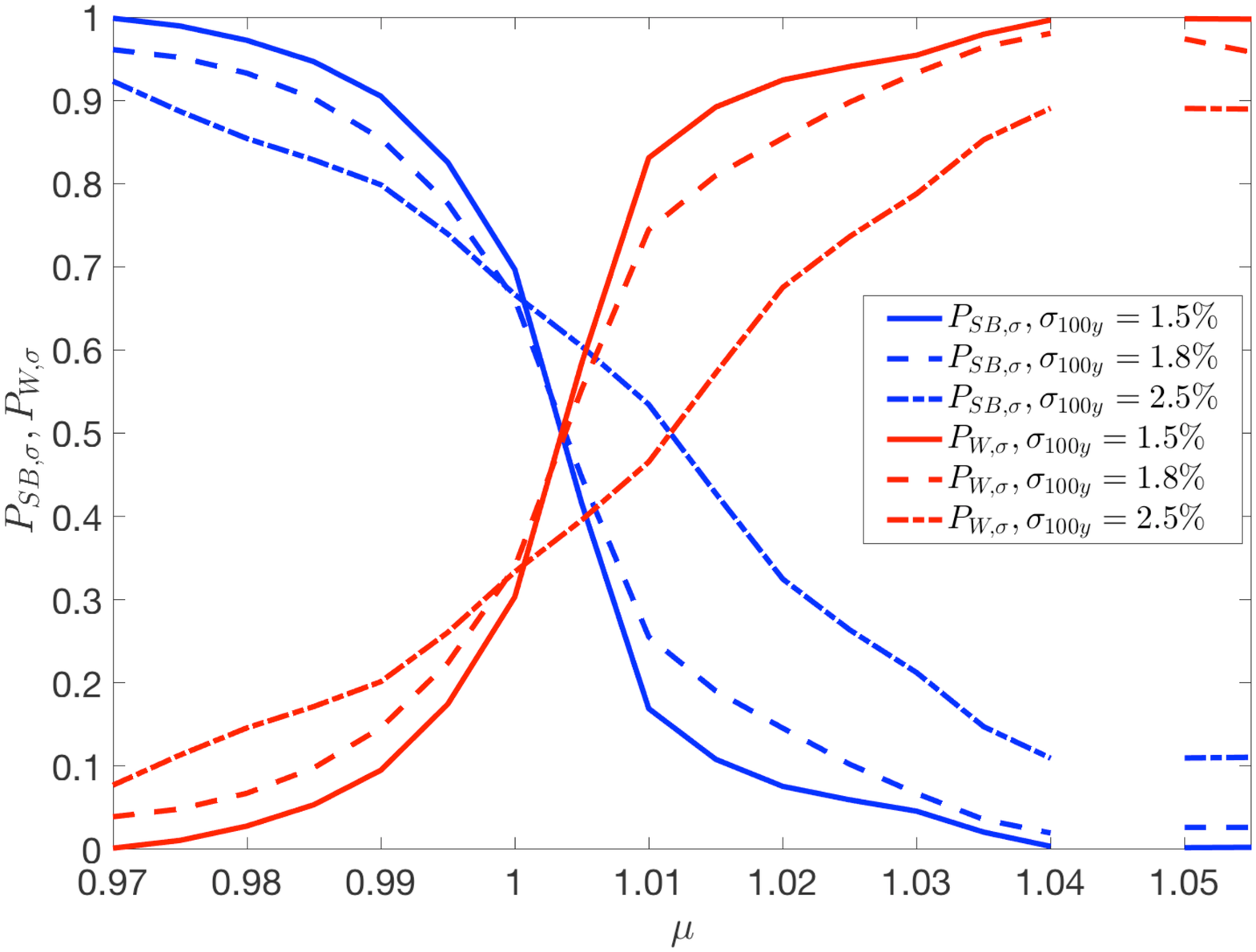}%
\caption{Projection of the measure on the variable $\langle T_S \rangle$ {\color{black}(units of $K$)} for $\sigma_{100y}=1.5\%$ (Panel a),  $\sigma_{100y}=1.8\%$ (Panel b), and  $\sigma_{100y}=2.5\%$ (Panel c). The spacing of the isolines is the same in the three panels. Panel d): Fraction of the measure supported in the basin of attraction of the SB state as a function of $\mu$ and of the noise intensity. As the noise decreases, we observe a fast transition between $SB$- and $W$-dominated  populations for $\mu=\mu_{crit}\approx1.005$, which corresponds to a first-order phase transition. \label{populations}}
\end{figure}

Equation \ref{eq:pop} indicates that, in the weak-noise limit, all of the measure will be concentrated on the attractor featuring the lowest value for the quasi-potential $\Phi$. Since for $\mu<S^*_{W\rightarrow SB}/S^*_0$ only the SB state is realised, physical intuition suggests that for low values of $\mu$ within the range of bistability, the SB attractor should contain all the mass in the limit of weak noise, as one can also anticipate by looking at Fig. \ref{mu098}. Conversely, one expects that for high values of  $\mu$ within the range of bistability, the $W$ should be dominant in the weak noise limit. It is reasonable (yet far from obvious or rigorous) to expect that there should be a critical value of $\mu=\mu_{crit}$ separating the two regimes. Following \cite{Lucarini2017}, we consider 18 equally spaced values of $\mu$ ($\Delta\mu=0.005$) within the multistable regime. For each of these values of  $\mu$ (excluding the case of $\mu=1.045$, where three stable states are realised) the fraction of the population residing within the basin of attraction of the deterministic W attractor $P_{W,\sigma}(\mu)$ and its complement, residing in the basin of attraction of the SB attractor $P_{SB,\sigma}(\mu)$. Related results are shown in Fig. \ref{populations}. We show how the probability distribution of the variable $\langle T_S \rangle$ depends on $\mu$ for three different noise levels: $\sigma_{100y}=1.5\%$ (Panel a),  $\sigma_{100y}=1.8\%$ (Panel b), and  $\sigma_{100y}=2.5\%$ (Panel c). In these panels we superimpose the bifurcation diagram reported in Fig. \ref{oldbifurcation}a). We observe that as the noise is reduced, for all values of $\mu$ the distributions are a) more peaked around the attractors; and b) one of the attractors becomes clearly dominant. 

In Panel d) the plot for each value of $\mu$ the integral of the pdfs reported in the three panels a), b), c) up to the values of $\langle T_S \rangle$ corresponding to the M state (green continuous and dotted lines). To a very good degree of approximation, this corresponds to the integral of the invariant measure over the support of the deterministic basin of attraction of the $SB$ climate. We obtain that for decreasing values of the noise intensity, the emerging invariant measure converges to the  deterministic measure supported on the SB attractor for $\mu\leq\mu_{crit}\approx1.005$, while the invariant measure converges to the deterministic measure supported on the W attractor for  $\mu\geq\mu_{crit}\approx 1.005$. The absolute minimum of the quasi-potential $\Phi$ is realised in the W attractor for $\mu\geq\mu_{crit}$ and in the SB attractor for  $\mu\leq\mu_{crit}$. The changeover is, curiously, quite close to the reference case $\mu=1$, for which the weak-noise limit of the measure is given by the $SB$ state.  

We add a note on the uncertainty associated to the figures reported in Fig. \ref{populations}d). We remark that for all values of $\mu$ and $\sigma$ we have used simulations lasting at least $10^4$ $y$. For very low ($\leq0.98$) and very large ($\geq1.02$) values of $\mu$ in the considered range, the simulation length does not allow for observing more than few transitions for the two lowest considered noise levels. Therefore, according to the fact that the transitions occur following a Poisson law, one expects in this range an uncertainty on the figures reported in Fig. \ref{populations} of the order of the values of the smaller between $P_{SB,\sigma}(\mu)$ and $P_{W,\sigma}(\mu)$. This corresponds, in fact, to a low uncertainty, because most of the mass is concentrated near one of the two deterministic attractors. The uncertainty is quite small for $0.98\leq\mu\leq1.03$, because, in all cases, we observe relatively many transitions. The uncertainty in this range can be safely estimated to be below 5\%. Summarising, while the values reported in Fig. \ref{populations}d) might have some non-negligible uncertainties, it seems that the estimate of $\mu_{crit}$ is quite robust.  

Finally, we have verified that for all values of $\mu$ the escape time $\tau_\sigma^{W\rightarrow SB}$ and $\tau_\sigma^{SB\rightarrow W}$ grow rapidly with decreasing values of the intensity of the noise for all values of $\mu$. In agreement with what is shown in Fig. \ref{populations}, we have that $\tau_\sigma^{SB\rightarrow W}$ grows more rapidly than $\tau_\sigma^{W\rightarrow SB}$ for $\mu\leq\mu_{crit}$ (and viceversa for $\mu\geq\mu_{crit}$). 
As a result, in the weak-noise limit an individual trajectory might be trapped for a very long time in the metastable, non-asymptotic state. We remark that we have not  performed here a systematic evaluation of the exponential relationship between the escape times and $\sigma$, as instead done for $\mu=0.98$ and shown in Fig. \ref{mu098tau}, also because, as for the reasons explained before, this would require computational resources that are beyond what has been allocated  for this study. We remark that this would allow for evaluating for each value of $\mu$ the difference between the value of the quasi-potential realised at the attractors and  at the M state. The algorithm proposed in \cite{Bodai2018} can be very useful to reduce the computational burden. 

We can say that  for $\mu=\mu_{crit}$ our system exhibits a behaviour that is reminiscent of a first-order phase transition for near-equilibrium statistical mechanical systems, like  a liquid-gaseous transition. In our case, $\mu$ is the control parameter (corresponding to the temperature in the equilibrium case), the quasi-potential $\Phi$ is the equivalent of a thermodynamic potential, the scaling factor for the noise intensity $\sigma$ is the equivalent of (square root of) the temperature, and the globally averaged surface temperature $\langle T_S \rangle$ is the natural order parameter, e.g. density. The discontinuous change in the properties of the system for $\mu=\mu_{crit}$ is associated to the change in the amount of absorbed and emitted radiation, as a result of the macroscopic change in the albedo of the planet due to the discontinuity in the position of the ice-line. We remark that choosing a different noise law would in general lead to a different value of $\mu_{crit}$, as a result of the fact that the functional form of $\Phi$ would be different.

{\color{black}\subsection{Relevance of the Choice of the It\^o Convention for the Noise}\label{itovsstrato}
A reasonable question to ask concerns to what extent our results are sensitive to the fact that we have chosen the It\^o convention for the noise, which provides the starting point of the results discussed in Sect. \ref{stochastic} and \ref{escape}. We argue that choosing other conventions would not alter essentially our findings because, to a first approximation, the stochastic forcing we have introduced can be treated as corresponding to perturbing the system with additive noise of  different strengths near the cold and W attractors, plus a transition region between the two attractors (which is evidently very sparsely populated by the system), where the  effective intensity of the noise decreases with the globally averaged surface temperature $\langle T_S\rangle$ and the multiplicative nature of the noise is more evident.  

In the phase space region near the cold attractor, we have that $1-\alpha(\phi,T_S)\sim 0.4$ since the temperature $T_S$ is extremely low and the planet is fully glaciated (or almost entirely so), so that  $\alpha(\phi,T_S)$ is virtually constant,  with $\alpha(\phi,T_S)\sim\alpha_{min}$.  Near the W attractor, the properties of the field $\alpha(\phi,T_S)$ are slightly more complex, because part of the planet is glaciated and part of it is ice-free. Nonetheless, to a first approximation, 
the ratio of the variance of the noise in the SB attractor vs. W attractor is of the order $((1-\alpha_{min})/(1-\alpha_{W}))^2\sim 3$. Loosely speaking, the competing W and SB climate states have different statistical mechanical, microscopic - as well as  thermodynamical, macroscopic - temperatures.}

\section{An Alternative Construction of the  M States using Stochastic Perturbations}\label{appa}
As discussed above, the construction of saddles for multistable systems is far from being a trivial task, and requires the use of the edge tracking algorithm introduced in \cite{Skufca2006,Schneider2007} and used also by the authors in \cite{Bodai2014,Lucarini2017}. We  wish to provide here a proof of concept of an alternative procedure for constructing the saddles - especially relevant when they are complex M states - without resorting to such an algorithm, but rather using only direct numerical simulations. The procedure discussed below might be useful when the edge tracking algorithm is hard to implement. As an example, this could be the case when the presence of similar time scales associated to the instability along and across the basin boundary might hinder an accurate computation of the saddles. Alternatively, it can be seen as a way to test the results obtained from the study of the deterministic dynamics.

The idea is to exploit the fact that, as discussed in Section \ref{Mathematics}, under rather general conditions on the noise law, the saddle, in the weak noise limit, acts  as the gate for noise-induced transitions between the competing attractors. We then propose to proceed as follows. Let's consider the following stochastic differential equations: 
\begin{equation}\label{eqapp2}
{d{\mathbf{x}}(t)}=\mathbf{F}(\mathbf{x(t)})dt+\sigma\mathbf{s_k}(\mathbf{x})d\mathbf{W},\quad k=1,\ldots,K.
\end{equation}
We consider the possibility of perturbing the deterministic flow with $K$ different noise laws, defined by the $K$ functions $\mathbf{s_k}(\mathbf{x})${\color{black}, each leading to a noise with different covariance matrix $C_{ij,k}(\mathbf{x})$}. We assume, for simplicity, that the deterministic system defined by ${\dot{{\mathbf{x}}}(t)}=\mathbf{F}(\mathbf{x(t)})$ is bistable. We choose  $K$ noise laws such that they obey the hypotheses discussed in Sects. \ref{stochastic} and \ref{escape}. 

{\color{black}In the weak noise limit $\sigma\rightarrow 0$, the invariant measure of the $k^{th}$ SDE can be written as  $\Pi_{\sigma,k} \sim \exp\left(-\frac{2\Phi_k(\mathbf{x})}{\sigma^2}\right)$. Clearly, for each noise law the  quasi-potential $\Phi_k(\mathbf{x})$ is different. Nonetheless, as discussed above, in all cases $\Phi_k(\mathbf{x})$  has a local minimum (and is constant) on the support of the two deterministic attractors, and it is a saddle with constant value on the saddle separating the two attractors. We here have to assume that either the saddle is unique, or that all the quasi-potentials $\Phi_k(\mathbf{x})$ select the same saddle as the one with the lowest quasi-potential.

Additionally, for each of the $K$ SDEs the drift flow is split differently between the gradient-like component and the rest (see Eq. \ref{eq:decomposition}). Therefore, as suggested by Eq. \ref{eqappi}, the instantonic paths are also different; yet, they  connect the same attractors to the same saddle. }Assuming that the attractors and the saddles are points or at least small sets given in a coarse-grained description of the phase space, we can identify the saddle as the only point in space where the instantons  corresponding to all the $K$ noise laws will intersect. 

\begin{figure}
a) \includegraphics[trim=0cm 0.0cm 0cm 0cm, clip=true, width=0.45\textwidth]{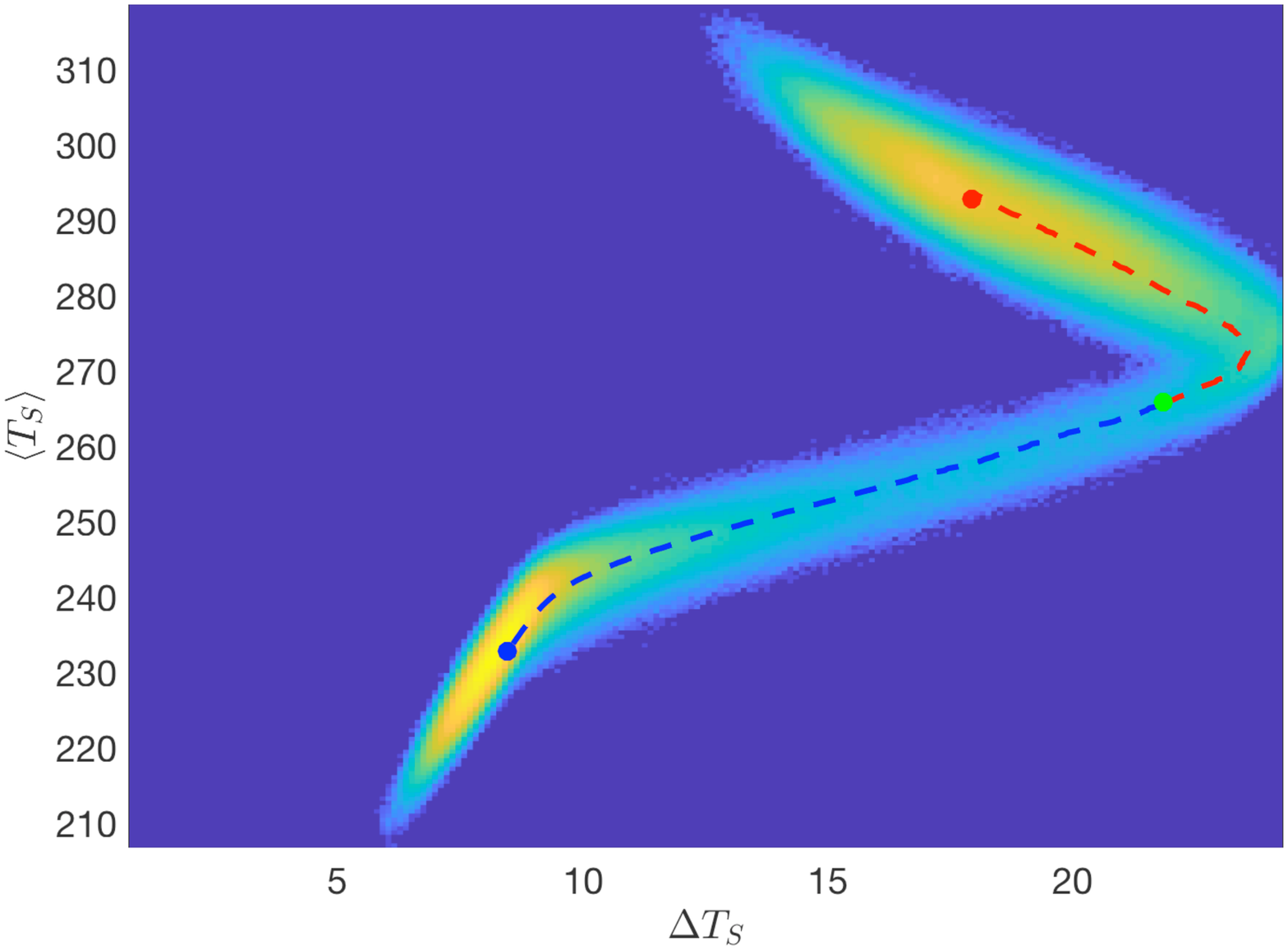} b) \includegraphics[trim=0cm 0.0cm 0cm 0cm, clip=true, width=0.45\textwidth]{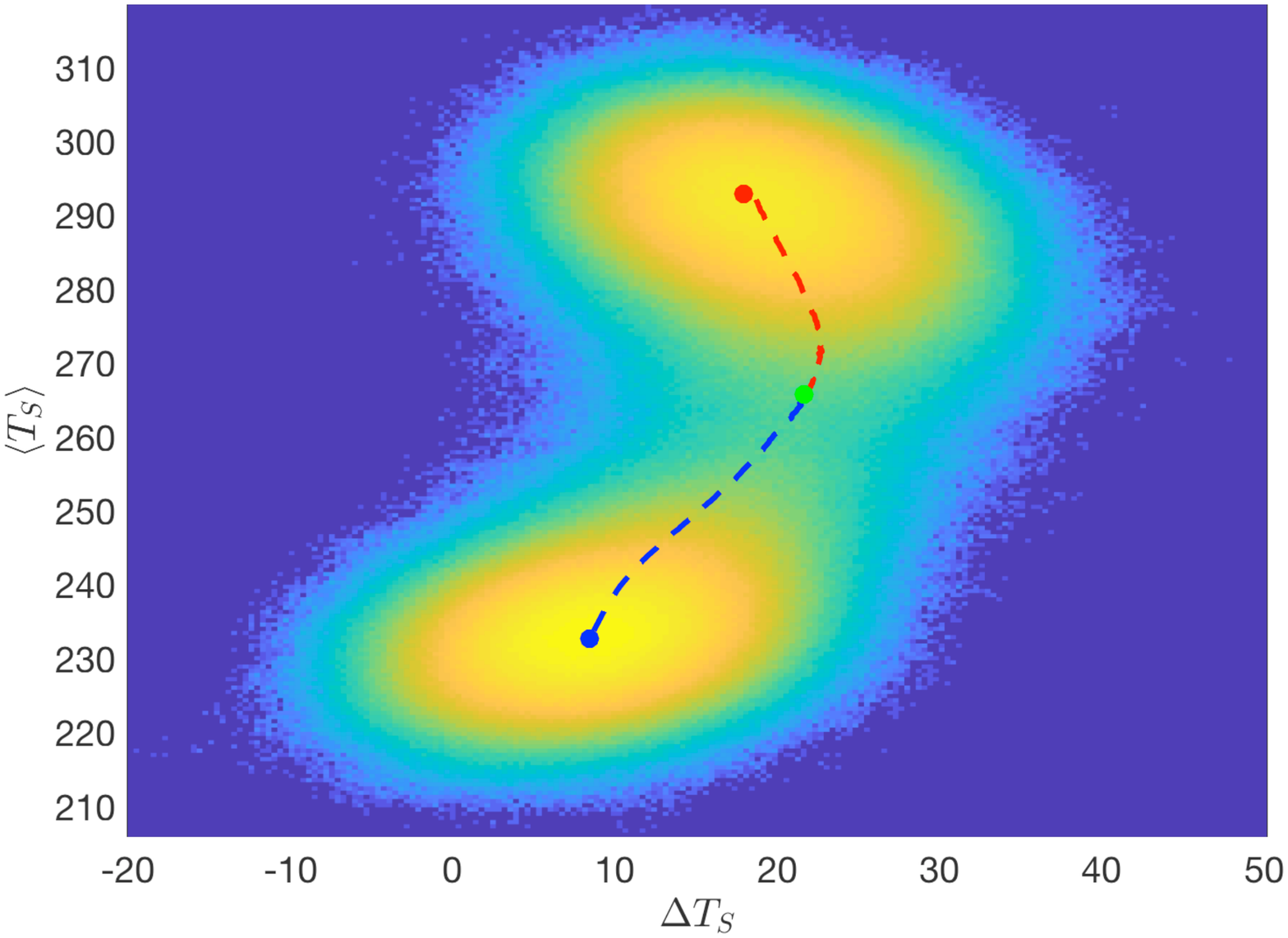}\\
c) \includegraphics[trim=0cm 0.0cm 0cm 0cm, clip=true, width=0.45\textwidth]{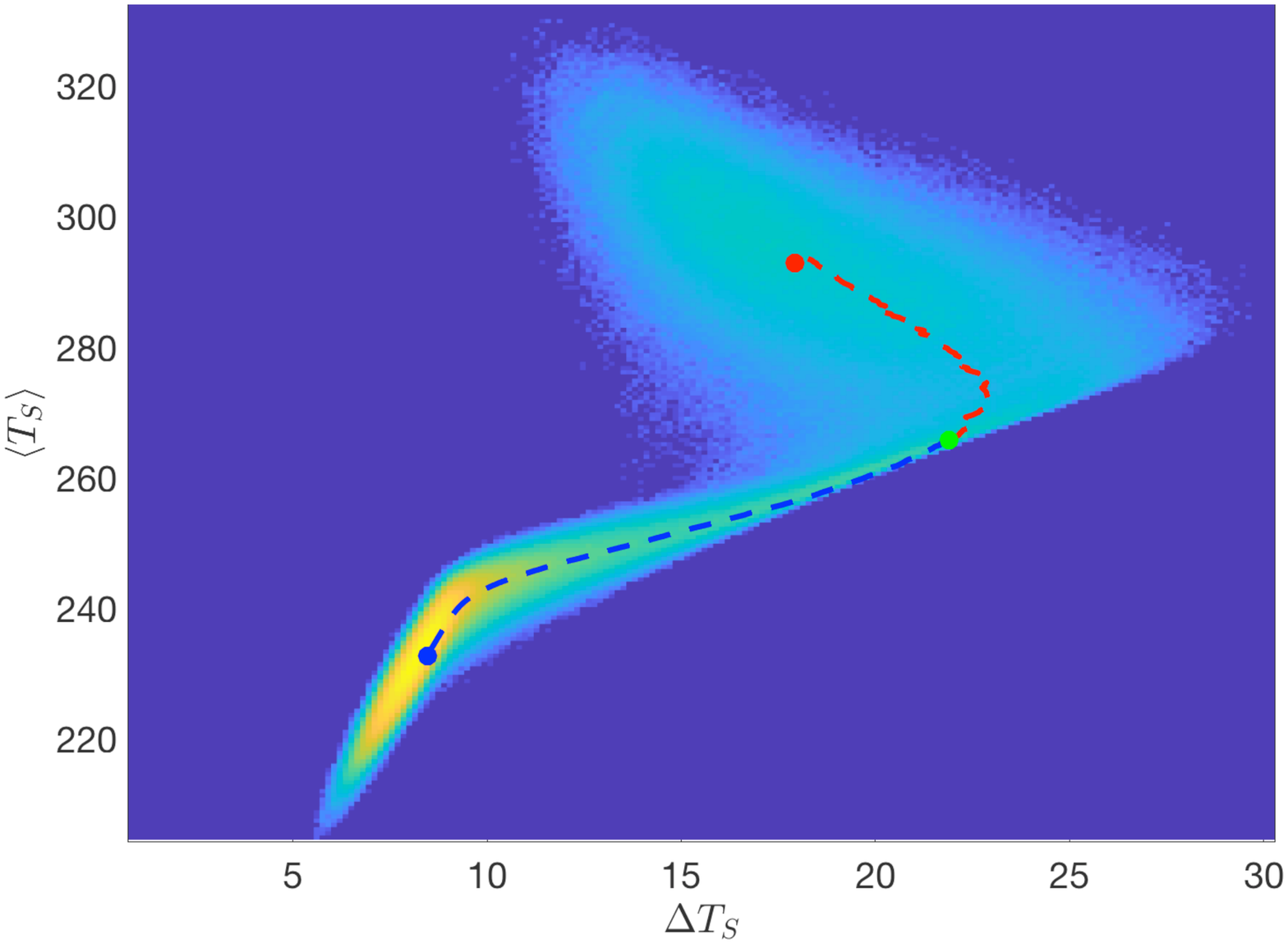} d) \includegraphics[trim=0cm 0.0cm 0cm 0cm, clip=true, width=0.45\textwidth]{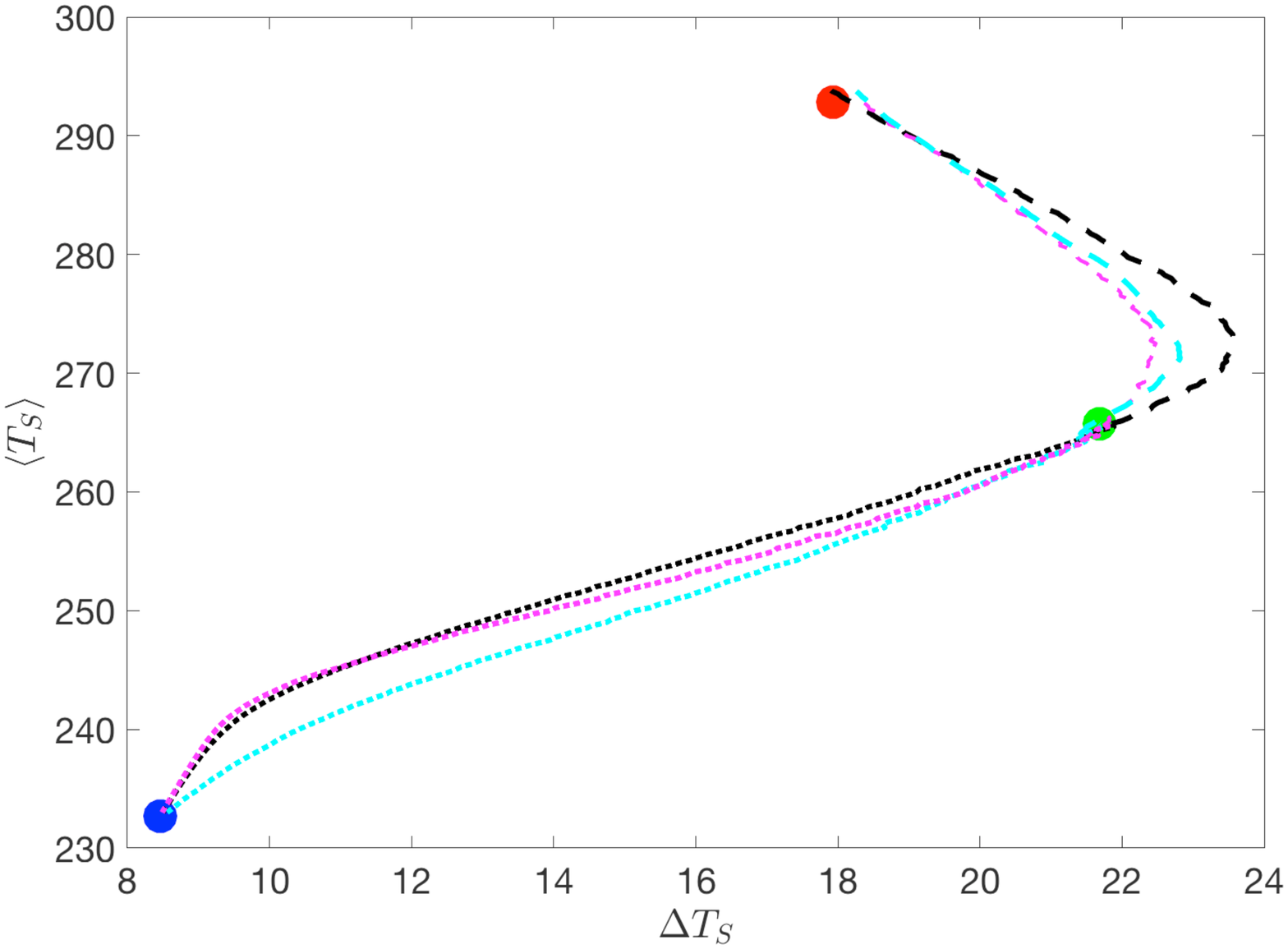}
\caption{Projected measure in the $(\Delta T_S,\langle T_S\rangle)$ space  {\color{black}(units of $K$ on both axes)} for selected stochastically perturbed simulations of the Ghil-Sellers models and related instantons. All results have been obtained with simulations lasting in $1.6\times10^6$ model years. a) Additive noise $s_1$: Logarithm of the pdf; $W\rightarrow SB$ instanton  (red dashed line); $SB\rightarrow W$ instanton (blue dashed line); W attractor (red dot); SB attractor (blue dot); saddle (green dot). The results have been obtained with $\sigma=0.4$. b) Same as in a), for additive noise $s_2$ (results obtained setting $\sigma=2$. c) Same as in a), for the multiplicative noise used in the rest of the paper (results obtained setting $\sigma=1.0$. d) The instantons from a) (black line), b) (cyan line) and c) (magenta line) are plotted together. Dashed (dotted) lines correspond to the $W\rightarrow SB$ ($SB\rightarrow W$) instantons. They cross at the saddle (green dot) and at the W attractor (red dot) and at the SB attractor (blue dot). }\label{figureappendix}
\end{figure}

In order to show that this approach does indeed work, we investigate the properties of $K=3$ variants of the Ghil-Sellers diffusive model we studied in \cite{Bodai2014}, differing for the law of the stochastic perturbation impacting the energy balance of the climate system. The Ghil-Sellers diffusive model can be written by removing the atmosphere-ocean coupling term in Eq. \ref{1DEBM}. In order to conform to Eq. \ref{eqapp} (note that we treat below the numerical discretization of a stochastically perturbed partial different equation), we  express the three stochastically perturbed models as follows:
\begin{equation}
\frac{\partial T_S(\phi,t)}{\partial t}= \mu\frac{S_0^*}{4}\frac{I(\phi)}{C(\phi)}(1-\alpha(\phi,T_S))-\frac{O(T_S)}{C(\phi)}-\frac{D_\phi[T_S]}{C(\phi)}+\sigma s_k(\phi,T_S) \frac{dW}{dt},\quad k=1,2,3,\label{1DEBM_mod}
\end{equation}
where $dW$ is the increment of a one-dimensional Wiener process, and we have:
\begin{align}
 s_1(\phi,T_S) &=  \mu \frac{S_0^*}{4}\frac{I(\phi)}{C(\phi)}, \label{1DEBM_mod1}\\
 s_2(\phi,T_S) &= 1  \label{1DEBM_mod2}, \\
 s_3(\phi,T_S) &=  \mu \frac{S_0^*}{4}\frac{I(\phi)}{C(\phi)},(1-\alpha(\phi,T_S)). \label{1DEBM_mod3}
\end{align}
Specifically, we have that  $s_1(\phi,T_S)= s_1(\phi)$ and $s_2(\phi,T_S)= s_2(\phi)$ correspond to additive noise laws, which feature different diffusion matrices. Instead, $s_3(\phi,T_S)$ is a multiplicative noise law as a result of the temperature-dependence of the albedo and is closely related to what was studied in the rest of the paper; see Eq. \ref{1DEBM}. We  construct for these three SDEs the invariant measure and, by stochastically averaging, we estimate the instantons connecting the attractors and the saddles. These sets are simple points. We make sure that the instantons are estimated using very weak noise amplitudes. Results are presented in Fig. \ref{figureappendix}, where we show the projections on the $(\Delta T_S,\langle T_S\rangle)$ space. Panels a), b), and c) show the invariant measures and the instantons constructed for the noise laws $s_1$ (using $\sigma=0.4$) , $s_2$ (using $\sigma=2.0$), and $s_3$ (using $\sigma=1.0$), respectively. Note that the instanton constructed with the multiplicative noise law looks qualitatively different from what is show in Fig. \ref{mu100_invariant}-\ref{instantonsmu100}, as a result of the lack of atmospheric motions in the simpler model discussed here. Panel d) portrays the instantons constructed for the three noise laws. Indeed, we have a confirmation that all of them are different, as a result of the different noise laws of the three SDEs, and intersect at the attractors and at the saddle. The position of the saddle in the projected phase space can be identified through this geometric procedure, {\color{black}which is based exclusively on direct numerical simulations.} Considering additional noise laws  can be helpful in resolving possible geometrical degeneracies due to the use of projections. Projecting in more than two dimensions could also serve a similar scope and provide a better understanding of the alternative transition paths.

\section{Conclusions}\label{Conclusions}

The goal of this paper has been the investigation of the properties of the noise-induced transitions across the multiple basins of attractions in an intermediate complexity climate model with $O(10^4)$ degrees of freedom, describing the coupled evolution of atmospheric (fast) and oceanic (slow) variables. The model features the co-esistence of W and SB attractors for a fairly broad range of values of the solar irradiance.  In a previous investigation we had been able to construct the full phase portrait of the deterministic version of the climate model considered here, and had constructed, beside the attractors, the M states of the climate system in the region of bistability \cite{Lucarini2017}. 

The stochastic forcing is introduced here as a random modulation for the incoming solar radiation, and leads to a nontrivial multiplicative noise law, because the radiative forcing is affected by the albedo of the surface, which, in turn, depends on the surface temperature. The noise, by allowing transitions between the deterministic basins of attraction, allows for establishing an (apparently) ergodic invariant measure of the system. The theory of stochastic differential equations indicates that for systems obeying the hypoellipticity condition, and for a suitable class of noise laws including additive laws and special multiplicative laws like the one considered here, one can write fairly generally the invariant measures in terms of a large deviation law, where the rate function can be identified with the quasi-potential. {\color{black}We have clarified how to compute the quasi-potential from the drift and volatility fields of the SDE and explained its property of being a Lyapunov function. The quasi-potential has local minima on the deterministic attractors, and has a saddle behaviour at the M states.}  Additionally, in the weak noise limit, transitions take place along special paths called instantons, which link the deterministic attractors and the M states. While, for a given deterministic dynamics, instantons corresponding to different noise laws follow different paths, they all link the same deterministic attractors to the same M states. {\color{black}We have shown how this property can be exploited to geometrically construct M states directly from direct numerical simulations of stochastic systems. Indeed, while the edge tracking algorithm applied to the deterministic system is \textit{a priori}  the preferred choice for finding  M states, it might become nontrivial to implement in complex numerical models where the intermediate states constructed by bisection might correspond to regions where the model is numerically unstable, possibly because the realised physical fields are extremely exotic or non-realisable.}

We have studied in detail the noise-induced transitions between the deterministic basins of attraction in the range of multistability, extending the results presented in a short communication \cite{LB2018}. We have shown that by studying how the average escape time depends on the intensity of the noise it is possible to estimate the difference between the value of the quasi-potential at the M state and at the attractor that the trajectories are escaping from. The estimates of the instantons are shown to become more precise as we consider simulations where weaker noise is considered. The instanton, in a case of special interest where the M state was shown to undergo a symmetry-break process, selects as optimal point of passage between the SB and the W climate the transient M state instead of the asymptotic one, possibly as a result of the finite amplitude of the noise. 

\begin{figure}
\includegraphics[trim=0.5cm 0.5cm 0.5cm 0.0cm, clip=true, width=0.5\textwidth]{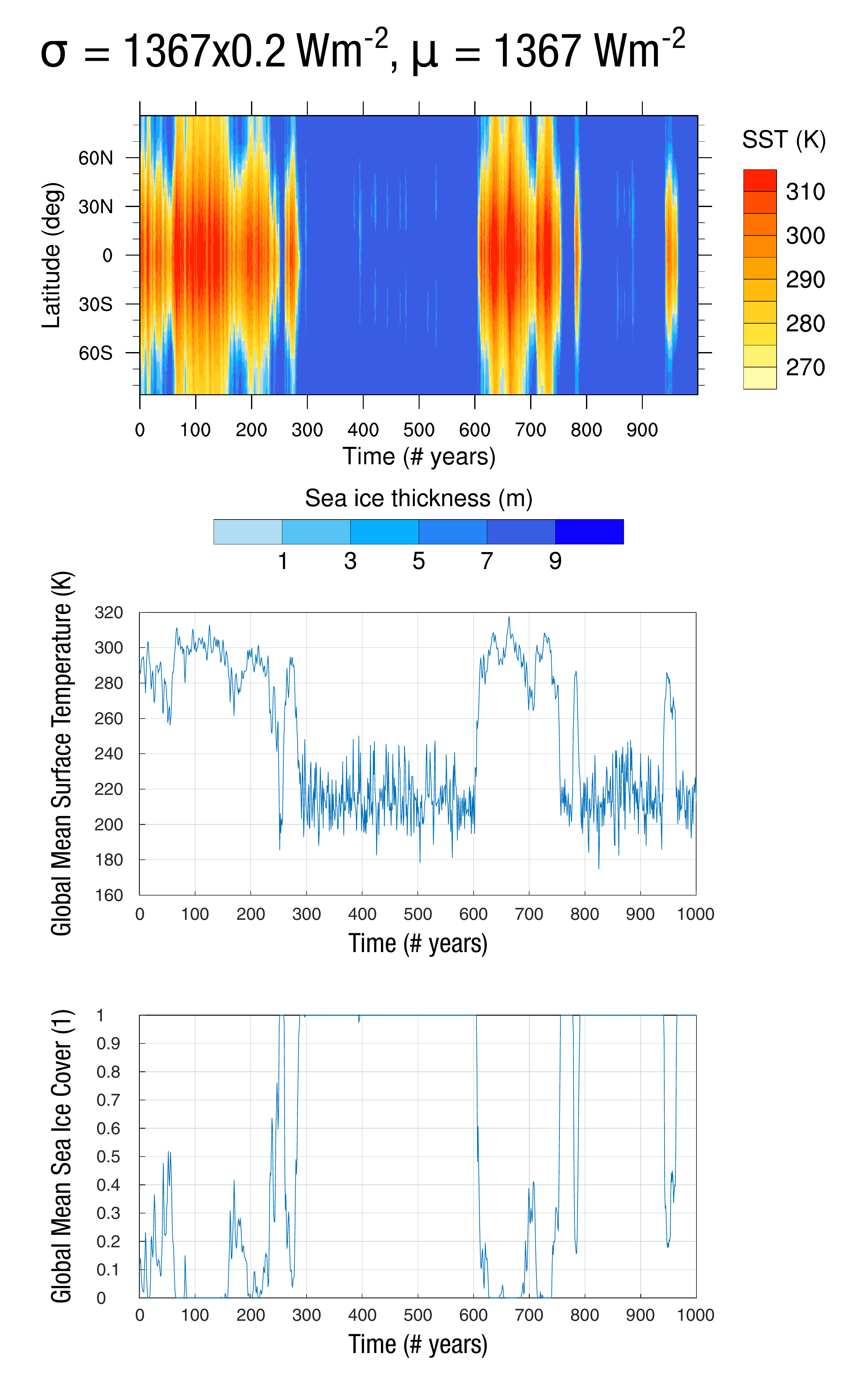}%
\caption{Example of noise-induced transitions between SB and W climate state for present-day solar constant ($\sigma_{100y}=2\%$) for the climate model PLASIM \cite{Fraedrich2005b} in a simulation of $10^3$ $y$. From top to bottom: zonal averages of the temperature field and of the sea ice cover; globally averaged surface temperature; fraction of ice-covered ocean. The characteristic escape times is comparable with what is obtained in the simpler model PUMA-GS used in this work (courtesy of {\color{black}F. Lunkeit}). \label{kilic}}
\end{figure}

Finally, by studying how the populations of the W and SB climate change as a function of the intensity of the noise, and using the large deviation law for the measure predicted by the theory, we find an estimate of a critical value of $\mu=\mu_{crit}\approx 1.005$ such that for $\mu\geq\mu_{crit}$ the zero-noise limit of the invariant measure is supported on the W deterministic attractor, while  for $\mu\leq\mu_{crit}$ the weak noise limit of the invariant measure is supported on the SB attractor. The asymptotic state corresponds to the attractor featuring the lowest value of the quasi-potential. 


These results obtained here indicate that, as soon as noise in some form is added to the system, multistability is factually lost in the weak-noise limit, as the noise law is responsible for selecting, for each value of the control parameter (here $\mu$), a specific asymptotic state. Changing the value of the control parameter, one will find one or more abrupt transitions in the statistical properties of the system (here realised at $\mu_{crit}$), i.e., in other terms discontinuities in the response of the system to changes in the control parameter. What happens in our model at $\mu=\mu_{crit}$ is mathematically analogous of a first-order phase transition occurring in a near-equilibrium statistical mechanical system. We remark that, since the escape time away from either attractor grows exponentially with the inverse  of the parameter controlling the variance of the noise, an individual trajectory might be trapped for very long time in a metastable state. 

In collaboration with  C. Kilic  (Bern) and F. Lunkeit (Hamburg) the authors have started some preliminary simulations where stochastic forcing is added to PLASIM \cite{Fraedrich2005b}, a much more complex climate model than the one used in this study (yet missing some essential ocean dynamical processes). As shown in Fig. \ref{kilic}, the first  results we have obtained are encouraging in indicating that the findings of this paper might be relevant for more realistic model configurations. For the future, we aim at obtaining detailed information on the large scale properties of the flow configurations leading to the noise-induced transitions, taking the thermodynamic lens we originally explored in \cite{Lucarini2010a}. 
An important question of specific relevance for paleoclimatic and planetary science studies is to understand whether the third climatic state found in \cite{Lucarini2017} for $\mu=1.045$ is recovered also for more realistic model configurations. {\color{black}Additionally, the complexity of the dynamical landscape of the climate system discussed in \cite{brunetti2019} suggests the existence of a possibly topologically non-trivial network of transition paths between the many competing attractors, each crossing an M state. Maybe the itinerancy between possibly many competing attractors might be a way to explaining the ultralow frequency variability of the climate. The computational needs of a naive approach to these issues seem prohibitive, so that one should definitely take advantage of rare events algorithms to construct instantonic trajectories \cite{rubino2009rare,Grafke_2013,Ragone2017}.}

The approach presented here,  based upon  combining the knowledge of the dynamical landscape of a multistable deterministic dynamical system with the analysis of the impacts of stochastic perturbations, seems of more general interested than the specific problem we have studied.  {\color{black}Along these lines, in Appendix \ref{appb} we speculate on the  the possible relevance of M states in the context of the theory of biological evolution and of synthetic models of evolution. Specifically, the idea is that the presence of qualitatively diverse historical paths of evolution might result from the the presence of M states in suitable defined dynamical landscapes, exactly because M states hinder predictability of the second kind in the sense of Lorenz.}

In the specific case of geosciences, we believe that it can be key for addressing the challenge of understanding tipping points in the Earth system \cite{Lenton2008} as well as providing insights to a large class of multistable systems \cite{Feudel2018}. We will dedicate future efforts exactly to exploring these research lines, in particularly looking at the Atlantic meridional overturning circulation, an element of the global ocean circulation that is well-know to have more than one competing \textit{modes of operation}. The occurrence of one or of the other state has important implications on the global climate, and rather dramatic ones for the regional climate of the North Atlantic sector; see \cite{Kuhlbrodt2007} for a review of this topic. Multistability has been recently reported in the von Karman turbulent flow in \cite{Faranda2017}: the methods proposed in this paper could elucidate paths and mechanisms underlying the transitions between the asymptotic states.

\section*{Acknowledgments}
VL and TB acknowledge the support received by the EU Horizon2020 projects Blue-Action (grant No. 727852) and CRESCENDO (grant No. 641816). VL acknowledges the support the EU Horizon2020 project TiPES (grant No. 820970), of the DFG SFB/Transregio project TRR181, and of the Royal Society (grant No. IEC/R2/170001). TB acknowledges the support received from the Institute for Basic Science (IBS), Republic of Korea, under IBS-R028-D1. The authors wish to thank T. T\'el for inspiring exchanges. {\color{black}VL wishes to thank S. Brusatte for suggesting some fascinating literature on evolutionary biology;} {\color{black} M. Ghil and} J. Yorke for great encouragement in the course of several cheerful discussions; T. Kuna, T. Lelievre, X.-M. Li, G. Pavliotis, and N. Zagli for providing insights on stochastic dynamics. This paper has been partly prepared during the Advanced Workshop on \textit{Nonequilibrium Systems in Physics, Geosciences, and Life Sciences} held at ICTP, Trieste, on May 15-24 2018 {\color{black} and during the Scientific Programme on the \textit{Mathematics of Climate and the Environment} held at the Institut H. Poincar\'e, Paris on September-December 2019.} The authors wish to thank {\color{black}F. Lunkeit} for kindly sharing Fig. \ref{kilic}. The authors wish to dedicate this paper to the memory of Bruno Eckhardt (1960-2019). 

\appendix
\section{An Interdisciplinary Outlook: Evolutionary Biology}\label{appb}
{\color{black}The concepts discussed in this paper might provide a conceptual and mathematical framework of possible interest for thinking at evolutive processes in biology and for constructing synthetic models of evolution. In the speculative discussion below, specifically, we want to highlight the role of M states in making possible the existence of multiple possible yet vastly different paths of historical development of biological systems.}

In the influential book discussing the Cambrian fossils from the Burgess shale and their importance for explaining the mechanisms of evolution, Gould \cite{G1989} presents some ideas on the scientific methodology inherent to historical sciences, and, specifically, to evolutionary biology. He clarifies that historical scientific explanations take the form of a narrative, whereby subsequent phenomena follow in a specific order\footnote{The related  \textit{storyline} approach is being currently proposed for studying weather and climate phenomena \cite{Shepherd2018}.}: the time-ordering and causal links between such phenomena can be discovered and convincingly explained when multiple, independent sources provide indication for the same historical pattern of change. Gould argues that this  method is fundamentally different from the classic - which he calls \textit{stereotypical} - scientific method \textit{\'a la Galileo} whereby one specific experiment can be repeated in idealised conditions, and the final outcomes of the experiment can be predicted using the fundamental laws of nature. The latter is the - indeed too simplified - version he gives of how \textit{hard sciences} work. An obvious difference between the two methods comes from the fact that certain experiments - like the evolution of the climate and of the biosphere of a specific planet - cannot be repeated. Another difference comes from the fact - Gould argues - that historical processes are dominated by \textit{contingency}: while we observe a specific path of historical development, many others of such paths are indeed compatible with the laws of nature, and could have been realised had the system been forced in a slightly different way in the past. We are able to understand the mechanisms of historical development, but not to predict accurately the specific path. Gould argues that the evolution in our planet could have led to fundamentally different forms of life, had the contingencies been  different in the distant past. The existence of our own species is the - a priori very unlikely - result of such contingencies. In general, in systems dominated by contingency, \textit{if we could run again the movie the outcome would be vastly different}. Gould's views on evolution have been criticised by authors, as Conway Morris, proposing that \textit{convergence}, rather than contingency, is the main mechanism of evolution, so that evolution is seen as a - mostly - deterministic path of change, of which nowadays we see the unavoidable outcome \cite{Morris1998}; see the debate in \cite{MorrisGould1998}. Some authors have proposed that both mechanisms are indeed in action, yet dominant at different scales of diversification of the organisms  \cite{Losos2017}.

We argue that  Gould's view can be put in the context of the mathematical framework discussed in this paper. Stochastically forced complex systems evolve in a phase space where an individual trajectory corresponds to a historical realisation of the system. The historical realisations, even starting from the same initial condition, can be vastly - even qualitatively - different if the deterministic dynamics supports the existence of multiple attractors, because different realisations can be trapped for very long times in very distant regions of the phase space. As discussed here, stochastic forcing allows for the system to jump across the various basins of attraction,  and the mechanism defining the evolution of the trajectory are defined by the differential equations. The predictability of the system, both of the first and second kind in the sense of Lorenz, is finite but non-zero. Finally, we can interpret the M states as the true agents of the contingency discussed by Gould. It is not the presence of a forcing, however strong, that changes radically the future history of the system, but rather the existence of special regions of the phase space - near the M states - where even small perturbations can force two nearby trajectories towards qualitatively different future histories. What  Conway Morris proposes is associated with a scenario where M states are either absent - so that the system is not multistable - or the noise is so weak that the probability of getting close to  an M  state is exceedingly low, despite the presence of stochastic perturbations, the system will be (almost) always quite close to a specific deterministic attractor, in correspondence with which basin of attraction the initial condition belongs to. In other terms. even \textit{if we could run again the movie}, i.e., run another simulation, the outcome would be very similar.

{\color{black}The interpretation given above receives some supports from recent results obtained on synthetic models of evolution. Using the the Tangled Nature model, which is conjectured to be a prototypical example of evolution, evolutive processes are interpreted as orbits of stochastic systems in a complex dynamical landscape featuring two or more competing metastable states \cite{Jones2010,Jensen2018}. This viewpoint, which has been openly inspired by Gould's ideas, is in close correspondence with what has been discussed in this paper. While in these works the language and methodology are eminently of statistical mechanical nature and they aim at detecting and studying the metastables states, the viewpoint proposed in this paper has a stronger emphasis  on understanding the transitions paths between such competing attractors through the M states.}


\begin{thebibliography}{100}

\bibitem{Budyko1969}
M.I. Budyko.
\newblock The effect of solar radiation variations on the climate of the earth.
\newblock {\em Tellus}, 21:611--619, 1969.

\bibitem{Sellers1969}
WD~Sellers.
\newblock A global climatic model based on the energy balance of the
  earthÐatmosphere system.
\newblock {\em J. Appl. Meteorol.}, 8:392--400, 1969.

\bibitem{Ghil1976}
M.~Ghil.
\newblock Climate stability for a {Sellers}-type model.
\newblock {\em J. Atmos. Sci.}, 33:3--20, 1976.

\bibitem{Kirschvink}
J.~L. Kirschvink.
\newblock {Late Proterozoic low-latitude global glaciation: The snowball
  Earth}.
\newblock In J.~W. Schopf and C.~Klein, editors, {\em The Proterozoic
  Biosphere: A Multidisciplinary Study}, chapter~8, pages 91--92. Cambridge
  University Press, 1992.

\bibitem{Hoffman2002}
Paul~F. Hoffman and Daniel~P. Schrag.
\newblock {The snowball Earth hypothesis: testing the limits of global change}.
\newblock {\em Terra Nova}, 14(3):129--155, 2002.

\bibitem{Pierrehumbert2011a}
R.T. Pierrehumbert, D.S. Abbot, Aiko Voigt, and D.~Koll.
\newblock {Climate of the Neoproterozoic}.
\newblock {\em Annual Review of Earth and Planetary Sciences}, 39(1):417--460,
  2011.

\bibitem{Hyde2000}
William~T. Hyde, Thomas~J. Crowley, Steven~K. Baum, and W.~Richard Peltier.
\newblock Neoproterozoic Ôsnowball earthÕ simulations with a coupled
  climate/ice-sheet model.
\newblock {\em Nature}, 405:425--429, 2000.

\bibitem{Voigt2010}
Aiko Voigt and Jochem Marotzke.
\newblock {The transition from the present-day climate to a modern Snowball
  Earth}.
\newblock {\em Climate Dynamics}, 35(5):887--905, 2010.

\bibitem{Lucarini2010a}
Valerio Lucarini, Klaus Fraedrich, and Frank Lunkeit.
\newblock {Thermodynamic analysis of snowball earth hysteresis experiment:
  Efficiency, entropy production and irreversibility}.
\newblock {\em Quarterly Journal of the Royal Meteorological Society},
  136(646):2--11, 2010.

\bibitem{Boschi2013}
R.~Boschi, V.~Lucarini, and S.~Pascale.
\newblock Bistability of the climate around the habitable zone: a thermodynamic
  investigation.
\newblock {\em Icarus}, 227:1724--1742, 2013.

\bibitem{crowley2001}
Thomas~J. Crowley, William~T. Hyde, and W.~Richard Peltier.
\newblock Co2 levels required for deglaciation of a Ònear-snowballÓ earth.
\newblock {\em Geophysical Research Letters}, 28(2):283--286, 2001.

\bibitem{Linsenmeier2015}
Manuel Linsenmeier, Salvatore Pascale, and Valerio Lucarini.
\newblock Climate of earth-like planets with high obliquity and eccentric
  orbits: Implications for habitability conditions.
\newblock {\em Planetary and Space Science}, 105:43 -- 59, 2015.

\bibitem{Kilic2017}
C.~Kilic, C.~C. Raible, and T.~F. Stocker.
\newblock Multiple climate states of habitable exoplanets: The role of
  obliquity and irradiance.
\newblock {\em The Astrophysical Journal}, 844(2):147, 2017.

\bibitem{Kilic2018}
Cevahir Kilic, Frank Lunkeit, Christoph~C. Raible, and Thomas~F. Stocker.
\newblock Stable equatorial ice belts at high obliquity in a coupled
  atmosphere{\textendash}ocean model.
\newblock {\em The Astrophysical Journal}, 864(2):106, sep 2018.

\bibitem{LucAstr2013}
V~Lucarini, S~Pascale, R~Boschi, E~Kirk, and N~Iro.
\newblock Habitability and multistablility in earth-like plantets.
\newblock {\em Astr. Nach.}, 334(6):576--588, 2013.

\bibitem{Abbot2018}
Dorian~S. Abbot, Jonah Bloch-Johnson, Jade Checlair, Navah~X. Farahat, R.~J.
  Graham, David Plotkin, Predrag Popovic, and Francisco Spaulding-Astudillo.
\newblock Decrease in hysteresis of planetary climate for planets with long
  solar days.
\newblock {\em The Astrophysical Journal}, 854(1):3, feb 2018.

\bibitem{Checlair2017}
Jade Checlair, Kristen Menou, and Dorian~S. Abbot.
\newblock No snowball on habitable tidally locked planets.
\newblock {\em The Astrophysical Journal}, 845(2):132, 2017.

\bibitem{lewis2007}
J.~P. Lewis, A.~J. Weaver, and M.~Eby.
\newblock Snowball versus slushball earth: Dynamic versus nondynamic sea ice?
\newblock {\em Journal of Geophysical Research: Oceans}, 112(C11):C11014, 2007.

\bibitem{abbott2011}
Dorian~S. Abbot, Aiko Voigt, and Daniel Koll.
\newblock The jormungand global climate state and implications for
  neoproterozoic glaciations.
\newblock {\em Journal of Geophysical Research: Atmospheres}, 116(D18), 2011.

\bibitem{Lucarini2017}
Valerio Lucarini and Tam\'as B\'odai.
\newblock Edge states in the climate system: exploring global instabilities and
  critical transitions.
\newblock {\em Nonlinearity}, 30(7):R32, 2017.

\bibitem{brunetti2019}
M.~Brunetti, J.~Kasparian, and C.~V{\'e}rard.
\newblock Co-existing climate attractors in a coupled aquaplanet.
\newblock {\em Climate Dynamics}, 53(9):6293--6308, 2019.

\bibitem{GomezLeal2018a}
Illeana G{\'{o}}mez-Leal, Lisa Kaltenegger, Valerio Lucarini, and Frank
  Lunkeit.
\newblock Climate sensitivity to carbon dioxide and the moist greenhouse
  threshold of earth-like planets under an increasing solar forcing.
\newblock {\em The Astrophysical Journal}, 869(2):129, dec 2018.

\bibitem{GomezLeal2018b}
Illeana G\'omez-Leal, Lisa Kaltenegger, Valerio Lucarini, and Frank Lunkeit.
\newblock Climate sensitivity to ozone and its relevance on the habitability of
  earth-like planets.
\newblock {\em Icarus}, 321:608 -- 618, 2019.

\bibitem{Kastings1993}
James~F. Kasting, Daniel~P. Whitmire, and Ray~T. Reynolds.
\newblock Habitable zones around main sequence stars.
\newblock {\em Icarus}, 101(1):108 -- 128, 1993.

\bibitem{Baladi2000}
Viviane Baladi.
\newblock {\em {Positive Transfer Operators and Decay of Correlations}}.
\newblock World Scientific, Singapore, 2000.

\bibitem{Pollicott1985}
Mark Pollicott.
\newblock {On the rate of mixing of Axiom A flows}.
\newblock {\em Inventiones Mathematicae}, 81(3):413--426, October 1985.

\bibitem{Ruelle1986}
David Ruelle.
\newblock {Resonances of Chaotic Dynamical Systems}.
\newblock {\em Physical review letters}, 56(5):405--407, 1986.

\bibitem{Tantet2018}
Alexis Tantet, Valerio Lucarini, Frank Lunkeit, and Henk~A Dijkstra.
\newblock Crisis of the chaotic attractor of a climate model: a transfer
  operator approach.
\newblock {\em Nonlinearity}, 31(5):2221, 2018.

\bibitem{Shiino1987}
Masatoshi Shiino.
\newblock Dynamical behavior of stochastic systems of infinitely many coupled
  nonlinear oscillators exhibiting phase transitions of mean-field type: H
  theorem on asymptotic approach to equilibrium and critical slowing down of
  order-parameter fluctuations.
\newblock {\em Phys. Rev. A}, 36:2393--2412, Sep 1987.

\bibitem{Pavliotis2014}
G.A. Pavliotis.
\newblock {\em Stochastic Processes and Applications: Diffusion Processes, the
  Fokker-Planck and Langevin Equations}.
\newblock Texts in Applied Mathematics. Springer New York, 2014.

\bibitem{Ragone2016}
F.~Ragone, V.~Lucarini, and F.~Lunkeit.
\newblock A new framework for climate sensitivity and prediction: a modelling
  perspective.
\newblock {\em Climate Dynamics}, 46(5):1459--1471, March 2016.

\bibitem{Lucarini2017b}
V.~Lucarini, F.~Ragone, and F.~Lunkeit.
\newblock Predicting climate change using response theory: Global averages and
  spatial patterns.
\newblock {\em Journal of Statistical Physics}, 166(3):1036--1064, February
  2017.

\bibitem{Lembo2019}
Valerio {Lembo}, Valerio {Lucarini}, and Francesco {Ragone}.
\newblock {Beyond Forcing Scenarios: Predicting Climate Change through Response
  Operators in a Coupled General Circulation Model}.
\newblock {\em arXiv e-prints}, page arXiv:1912.03996, Dec 2019.

\bibitem{Ruelle2009}
David Ruelle.
\newblock {A review of linear response theory for general differentiable
  dynamical systems}.
\newblock {\em Nonlinearity}, 22:855--870, 2009.

\bibitem{Ghil2008}
Michael Ghil, Micka\"{e}l~David Chekroun, and Eric Simonnet.
\newblock {Climate dynamics and fluid mechanics: Natural variability and
  related uncertainties}.
\newblock {\em Physica D: Nonlinear Phenomena}, 237(14-17):2111--2126, August
  2008.

\bibitem{Chekroun2011}
Micka\"{e}l~David Chekroun, Eric Simonnet, and Michael Ghil.
\newblock {Stochastic climate dynamics: Random attractors and time-dependent
  invariant measures}.
\newblock {\em Physica D: Nonlinear Phenomena}, 240(21):1685--1700, October
  2011.

\bibitem{CLR13}
A.~N. Carvalho, J.~Langa, and J.~C. Robinson.
\newblock The pullback attractor.
\newblock In {\em Attractors for infinite-dimensional non-autonomous dynamical
  systems}, volume 182 of {\em Applied Mathematical Sciences}, pages 3--22.
  Springer New York, 2013.

\bibitem{Ott.ea.1990}
F.~J. Romeiras, C.~Grebogi, and E.~Ott.
\newblock Multifractal properties of snapshot attractors of random maps.
\newblock {\em Phys. Rev. A}, 41(2):784, 1990.

\bibitem{DBT15}
G.~Dr{\'o}tos, T.~B{\'o}dai, and T.~T{\'e}l.
\newblock Probabilistic concepts in a changing climate: A snapshot attractor
  picture.
\newblock {\em Journal of Climate}, 28(8):3275--3288, 2015.

\bibitem{Dembo2010}
Amir Dembo and Jean-Dominique Deuschel.
\newblock Markovian perturbation, response and fluctuation dissipation theorem.
\newblock {\em Ann. Inst. H. PoincarŽ Probab. Statist.}, 46(3):822--852, 08
  2010.

\bibitem{Assaraf2018}
Roland Assaraf, Benjamin Jourdain, Tony Leli{\`e}vre, and Rapha{\"e}l Roux.
\newblock Computation of sensitivities for the invariant measure of a parameter
  dependent diffusion.
\newblock {\em Stochastics and Partial Differential Equations: Analysis and
  Computations}, 6(2):125--183, Jun 2018.

\bibitem{Lucarini2016}
Valerio Lucarini.
\newblock Response operators for markov processes in a finite state space:
  Radius of convergence and link to the response theory for axiom a systems.
\newblock {\em Journal of Statistical Physics}, 162(2):312--333, Jan 2016.

\bibitem{Chekroun2014}
Micka\"{e}l~David Chekroun, J.~David Neelin, Dmitri Kondrashov, J.~C.
  McWilliams, and Michael Ghil.
\newblock {Rough parameter dependence in climate models and the role of
  Ruelle-Pollicott resonances.}
\newblock {\em Proceedings of the National Academy of Sciences of the United
  States of America}, 111(5):1684--1690, March 2014.

\bibitem{Ghil2019}
Michael {Ghil} and Valerio {Lucarini}.
\newblock {The Physics of Climate Variability and Climate Change}.
\newblock {\em arXiv e-prints}, page arXiv:1910.00583, Oct 2019.

\bibitem{Bodai2014}
Tam{\'a}s B{\'o}dai, Valerio Lucarini, Frank Lunkeit, and Robert Boschi.
\newblock Global instability in the ghil--sellers model.
\newblock {\em Clim. Dyn.}, 44(11):3361--3381, 2014.

\bibitem{Grebogi1983}
Celso Grebogi, Edward Ott, and James~A. Yorke.
\newblock Fractal basin boundaries, long-lived chaotic transients, and
  unstable-unstable pair bifurcation.
\newblock {\em Phys. Rev. Lett.}, 50:935--938, Mar 1983.

\bibitem{Robert2000}
Carl Robert, Kathleen~T. Alligood, Edward Ott, and James~A. Yorke.
\newblock Explosions of chaotic sets.
\newblock {\em Physica D: Nonlinear Phenomena}, 144(1):44 -- 61, 2000.

\bibitem{Ott2002}
E.~Ott.
\newblock {\em Chaos in Dynamical Systems}.
\newblock Cambridge University Press, 2002.

\bibitem{LT:2011}
Y.-C. Lai and T.~T\'el.
\newblock {\em Transient Chaos}.
\newblock Springer, New York, 2011.

\bibitem{Skufca2006}
Joseph~D. Skufca, James~A. Yorke, and Bruno Eckhardt.
\newblock Edge of chaos in a parallel shear flow.
\newblock {\em Phys. Rev. Lett.}, 96:174101, May 2006.

\bibitem{Schneider2007}
Tobias~M. Schneider, Bruno Eckhardt, and James~A. Yorke.
\newblock Turbulence transition and the edge of chaos in pipe flow.
\newblock {\em Phys. Rev. Lett.}, 99:034502, Jul 2007.

\bibitem{Sieber2013}
D.~A.~W. Barton and J.~Sieber.
\newblock Systematic experimental exploration of bifurcations with noninvasive
  control.
\newblock {\em Phys. Rev. E}, 87:052916, May 2013.

\bibitem{Sieber2014}
Jan Sieber, Oleh~E. Omel'chenko, and Matthias Wolfrum.
\newblock Controlling unstable chaos: Stabilizing chimera states by feedback.
\newblock {\em Phys. Rev. Lett.}, 112:054102, Feb 2014.

\bibitem{Abrams2004}
Daniel~M. Abrams and Steven~H. Strogatz.
\newblock Chimera states for coupled oscillators.
\newblock {\em Phys. Rev. Lett.}, 93:174102, Oct 2004.

\bibitem{Omelchenko2018Nonlinearity}
O.~E. Omel'chenko.
\newblock The mathematics behind chimera states.
\newblock {\em Nonlinearity}, 31(5):R121, 2018.

\bibitem{LB2018}
Valerio Lucarini and Tam\'as B\'odai.
\newblock Transitions across melancholia states in a climate model: Reconciling
  the deterministic and stochastic points of view.
\newblock {\em Phys. Rev. Lett.}, 122:158701, Apr 2019.

\bibitem{Bell2004}
Denis~R. Bell.
\newblock {\em Stochastic Differential Equations and Hypoelliptic Operators},
  pages 9--42.
\newblock Birkh{\"a}user Boston, Boston, MA, 2004.

\bibitem{Hanggi1986}
Peter Hanggi.
\newblock Escape from a metastable state.
\newblock {\em Journal of Statistical Physics}, 42(1):105--148, Jan 1986.

\bibitem{Kautz1987}
R.L. Kautz.
\newblock Activation energy for thermally induced escape from a basin of
  attraction.
\newblock {\em Physics Letters A}, 125(6):315 -- 319, 1987.

\bibitem{Grassberger1989}
P~Grassberger.
\newblock Noise-induced escape from attractors.
\newblock {\em Journal of Physics A: Mathematical and General}, 22(16):3283,
  1989.

\bibitem{Freidlin1984}
M.~I. Freidlin and A.D. Wentzell.
\newblock {\em Random Perturbations of Dynamical Systems}.
\newblock Springer, New York, 1984.

\bibitem{Graham1991}
R.~Graham, A.~Hamm, and T.~T\'el.
\newblock Nonequilibrium potentials for dynamical systems with fractal
  attractors or repellers.
\newblock {\em Phys. Rev. Lett.}, 66:3089--3092, Jun 1991.

\bibitem{Hamm1994}
A.~Hamm, T.~T\'el, and R.~Graham.
\newblock Noise-induced attractor explosions near tangent bifurcations.
\newblock {\em Physics Letters A}, 185(3):313 -- 320, 1994.

\bibitem{Kraut2002}
Suso Kraut and Ulrike Feudel.
\newblock Multistability, noise, and attractor hopping: The crucial role of
  chaotic saddles.
\newblock {\em Phys. Rev. E}, 66:015207, Jul 2002.

\bibitem{Beri2005}
S.~Beri, R.~Mannella, D.~G. Luchinsky, A.~N. Silchenko, and P.~V.~E.
  McClintock.
\newblock Solution of the boundary value problem for optimal escape in
  continuous stochastic systems and maps.
\newblock {\em Phys. Rev. E}, 72:036131, Sep 2005.

\bibitem{Bouchet2016}
Freddy Bouchet and Julien Reygner.
\newblock Generalisation of the eyring--kramers transition rate formula to
  irreversible diffusion processes.
\newblock {\em Annales Henri Poincar{\'e}}, 17(12):3499--3532, Dec 2016.

\bibitem{G1989}
S.~J. Gould.
\newblock {\em {Wonderful Life: The Burgess shale and the Nature of History}}.
\newblock W.W. Norton, New York, 1989.

\bibitem{Feudel2018}
Ulrike Feudel, Alexander~N. Pisarchik, and Kenneth Showalter.
\newblock Multistability and tipping: From mathematics and physics to climate
  and brain—minireview and preface to the focus issue.
\newblock {\em Chaos: An Interdisciplinary Journal of Nonlinear Science},
  28(3):033501, 2018.

\bibitem{Lenton2008}
Timothy~M. Lenton, Hermann Held, Elmar Kriegler, Jim~W. Hall, Wolfgang Lucht,
  Stefan Rahmstorf, and Hans~Joachim Schellnhuber.
\newblock Tipping elements in the earth{\textquoteright}s climate system.
\newblock {\em Proceedings of the National Academy of Sciences},
  105(6):1786--1793, 2008.

\bibitem{Vollmer2009}
Juergen Vollmer, Tobias~M Schneider, and Bruno Eckhardt.
\newblock Basin boundary, edge of chaos and edge state in a two-dimensional
  model.
\newblock {\em New Journal of Physics}, 11(1):013040, 2009.

\bibitem{BodaiLucarini2020}
{Tam{\'a}s B{\'o}dai} and Valerio Lucarini.
\newblock {Rough basin boundaries in high dimension: Can we classify them
  experimentally}.
\newblock {\em arXiv e-prints}, page arXiv:XX.YY, Jan 2020.

\bibitem{Lorenz1975}
E.~N. Lorenz.
\newblock The physical bases of climate and climate modelling. climate
  predictability.
\newblock In {\em GARP Publication Series}, pages 132--136. WMO, 1975.

\bibitem{Hairer2011}
Martin Hairer.
\newblock On malliavin's proof of h\:omander's theorem.
\newblock {\em Bulletin des Sciences Math\'ematiques}, 135(6):650 -- 666, 2011.
\newblock Special issue in memory of Paul Malliavin.

\bibitem{Gaspard2002}
Pierre Gaspard.
\newblock Trace formula for noisy flows.
\newblock {\em Journal of Statistical Physics}, 106(1):57--96, Jan 2002.

\bibitem{Nardini2016}
Freddy Bouchet, Krzysztof Gawedzki, and Cesare Nardini.
\newblock Perturbative calculation of quasi-potential in non-equilibrium
  diffusions: A mean-field example.
\newblock {\em Journal of Statistical Physics}, 163(5):1157--1210, Jun 2016.

\bibitem{Ao2004}
P~Ao.
\newblock Potential in stochastic differential equations: novel construction.
\newblock {\em Journal of Physics A: Mathematical and General}, 37(3):L25--L30,
  jan 2004.

\bibitem{Yin2006}
L~Yin and P~Ao.
\newblock Existence and construction of dynamical potential in nonequilibrium
  processes without detailed balance.
\newblock {\em Journal of Physics A: Mathematical and General},
  39(27):8593--8601, jun 2006.

\bibitem{Zhou2012}
Joseph~Xu Zhou, M.~D.~S. Aliyu, Erik Aurell, and Sui Huang.
\newblock Quasi-potential landscape in complex multi-stable systems.
\newblock {\em Journal of The Royal Society Interface}, 9(77):3539--3553, 2012.

\bibitem{Brackston2018}
R.~D. Brackston, A.~Wynn, and M.~P.~H. Stumpf.
\newblock Construction of quasipotentials for stochastic dynamical systems: An
  optimization approach.
\newblock {\em Phys. Rev. E}, 98:022136, Aug 2018.

\bibitem{Tang2017}
Ying Tang, Ruoshi Yuan, Gaowei Wang, Xiaomei Zhu, and Ping Ao.
\newblock Potential landscape of high dimensional nonlinear stochastic dynamics
  with large noise.
\newblock {\em Scientific Reports}, 7(1):15762, 2017.

\bibitem{Graham1987}
R.~Graham.
\newblock Macroscopic potentials, bifurcations and noise in dissipative
  systems.
\newblock In L.~Garrido, editor, {\em Fluctuations and Stochastic Phenomena in
  Condensed Matter}, pages 1--34, Berlin, Heidelberg, 1987. Springer Berlin
  Heidelberg.

\bibitem{Graham1986}
R.~Graham and T.~T\'el.
\newblock Nonequilibrium potential for coexisting attractors.
\newblock {\em Phys. Rev. A}, 33:1322--1337, Feb 1986.

\bibitem{Bovier2004}
Anton Bovier, Michael Eckhoff, Véronique Gayrard, and Markus Klein.
\newblock Metastability in reversible diffusion processes i. sharp asymptotics
  for capacities and exit times.
\newblock {\em Journal of the European Mathematical Society}, 6:399--424, 10
  2004.

\bibitem{Bodai2018}
Tam{\'a}s {B{\'o}dai}.
\newblock {An efficient algorithm to estimate the potential barrier height from
  noise-induced escape time data}.
\newblock {\em arXiv e-prints}, page arXiv:1808.06903, Aug 2018.

\bibitem{Lelievre2015}
T.~Leli{\`e}vre.
\newblock Accelerated dynamics: Mathematical foundations and algorithmic
  improvements.
\newblock {\em The European Physical Journal Special Topics},
  224(12):2429--2444, Sep 2015.

\bibitem{Gesu2019}
Giacomo~Di Ges{\`u}, Tony Leli{\`e}vre, Dorian~Le Peutrec, and Boris Nectoux.
\newblock Sharp asymptotics of the first exit point density.
\newblock {\em Annals of PDE}, 5(1):5, Mar 2019.

\bibitem{Lucarini2019}
Valerio Lucarini.
\newblock Stochastic resonance for nonequilibrium systems.
\newblock {\em Phys. Rev. E}, 100:062124, Dec 2019.

\bibitem{puma}
Thomas Frisius, Frank Lunkeit, Klaus Fraedrich, and Ian~N. James.
\newblock Storm-track organization and variability in a simplified atmospheric
  global circulation model.
\newblock {\em Quarterly Journal of the Royal Meteorological Society},
  124(548):1019--1043, 1998.

\bibitem{saltzman_dynamical}
Barry Saltzman.
\newblock {\em Dynamical Paleoclimatology}.
\newblock Academic Press, New York, 2001.

\bibitem{Peixoto1992}
Jos~P Peixoto and Abraham~H Oort.
\newblock {\em {Physics of Climate}}.
\newblock American Institute of Physics, 1992.

\bibitem{Lucarini2014b}
Valerio Lucarini, Richard Blender, Corentin Herbert, Francesco Ragone,
  Salvatore Pascale, and Jeroen Wouters.
\newblock {Mathematical and Physical Ideas for Climate Science}.
\newblock {\em Reviews of Geophysics}, 52(4):1 -- 51, 2014.

\bibitem{Holton}
James~R. Holton.
\newblock {\em {An introduction to dynamic meteorology}}.
\newblock International Geophysics Series. Elsevier Academic Press,,
  Burlington, MA, 4 edition, 2004.

\bibitem{ZinnJustin1996}
J.~Zinn-Justin.
\newblock {\em Quantum Field Theory and Critical Phenomena}.
\newblock Oxford University Press, Oxford, 1996.

\bibitem{rubino2009rare}
G.~Rubino and B.~Tuffin.
\newblock {\em Rare Event Simulation using Monte Carlo Methods}.
\newblock Wiley, 2009.

\bibitem{Ragone2017}
F.~Ragone, J.~Wouters, and F.~Bouchet.
\newblock Computation of extreme heat waves in climate models using a large
  deviation algorithm.
\newblock {\em Proceedings of the National Academy of Sciences}, 115(1):24--29,
  2018.

\bibitem{Grafke_2013}
Tobias Grafke, Rainer Grauer, and Tobias Schäfer.
\newblock Instanton filtering for the stochastic burgers equation.
\newblock {\em Journal of Physics A: Mathematical and Theoretical},
  46(6):062002, jan 2013.

\bibitem{Fraedrich2005b}
Klaus Fraedrich, Heiko Jansen, Edilbert Kirk, Ute Luksch, and Frank Lunkeit.
\newblock {The Planet Simulator: Towards a user friendly model}.
\newblock {\em Meteorologische Zeitschrift}, 14(3):299--304, 2005.

\bibitem{Kuhlbrodt2007}
T~Kuhlbrodt, A~Griesel, M~Montoya, A~Levermann, M~Hofmann, and S~Rahmstorf.
\newblock {On the driving processes of the Atlantic meridional overturning
  circulation}.
\newblock {\em Atlantic}, 45(2004):RG2001, 2007.

\bibitem{Faranda2017}
D.~Faranda, Y.~Sato, B.~Saint-Michel, C.~Wiertel, V.~Padilla, B.~Dubrulle, and
  F.~Daviaud.
\newblock Stochastic chaos in a turbulent swirling flow.
\newblock {\em Phys. Rev. Lett.}, 119:014502, Jul 2017.

\bibitem{Shepherd2018}
Theodore~G. Shepherd, Emily Boyd, Raphael~A. Calel, Sandra~C. Chapman, Suraje
  Dessai, Ioana~M. Dima-West, Hayley~J. Fowler, Rachel James, Douglas Maraun,
  Olivia Martius, Catherine~A. Senior, Adam~H. Sobel, David~A. Stainforth,
  Simon F.~B. Tett, Kevin~E. Trenberth, Bart J. J.~M. van~den Hurk, Nicholas~W.
  Watkins, Robert~L. Wilby, and Dimitri~A. Zenghelis.
\newblock Storylines: an alternative approach to representing uncertainty in
  physical aspects of climate change.
\newblock {\em Climatic Change}, 151(3):555--571, Dec 2018.

\bibitem{Morris1998}
S.~Conway Morris.
\newblock {\em {The crucible of creation: the Burgess Shale and the rise of
  animals}}.
\newblock Oxford University Press, Oxford, 1998.

\bibitem{MorrisGould1998}
Simon~Conway Morris and Stephen~Jay Gould.
\newblock Showdown on the burgess shale.
\newblock {\em Natural History magazine}, 107(10):48--55, 1998.

\bibitem{Losos2017}
J.~B. Losos.
\newblock {\em {Improbable destinies: Fate, chance, and the future of
  evolution}}.
\newblock Riverhead Books, New York, 2017.

\bibitem{Jones2010}
Dominic Jones, Henrik~Jeldtoft Jensen, and Paolo Sibani.
\newblock Tempo and mode of evolution in the tangled nature model.
\newblock {\em Phys. Rev. E}, 82:036121, Sep 2010.

\bibitem{Jensen2018}
Henrik~Jeldtoft Jensen.
\newblock Tangled nature: a model of emergent structure and temporal mode among
  co-evolving agents.
\newblock {\em European Journal of Physics}, 40(1):014005, dec 2018.

\end{thebibliography}
\end{document}